\documentclass[aps,pra,twocolumn,showpacs,preprintnumbers,amsmath,amssymb,footinbib]{revtex4-1}

\usepackage{amsmath}    % need for subequations
\usepackage{graphicx}   % need for figures
\usepackage{verbatim}   % useful for program listings
\usepackage{color}      % use if color is used in text
\usepackage{subfigure}  % use for side-by-side figures
\usepackage{hyperref}   % use for hypertext links, including those to external 
\newcommand{\Fig} [1]  {Fig.~\ref{#1}}
\newcommand{\Figs}[1]  {Figs.~\ref{#1}}

\newcommand{\gep} {GeP$_3$}
\newcommand{\bl} {(GeP$_3$)$_{\rm BL}$}
\newcommand{\blcr} {$\rm(GeP_3)_{BL}^{\rm Cr}$}
\newcommand{\qlcr} {$\rm(GeP_3)_{QL}^{\rm Cr}$}
\newcommand{\udud} {Cr$^{\uparrow\downarrow}$//Cr$^{\uparrow\downarrow}$}
\newcommand{\ud} {Cr$^{\uparrow\downarrow}$}
\newcommand{\uu} {Cr$^{\uparrow\uparrow}$}
\newcommand{\uuuu} {Cr$^{\uparrow\uparrow}$//Cr$^{\uparrow\uparrow}$}
\newcommand{\uudd} {Cr$^{\uparrow\uparrow}$//Cr$^{\downarrow\downarrow}$}
\newcommand{\ududud}{Cr$^{\uparrow\downarrow}$/Cr$^{\uparrow\downarrow}$/Cr$^{
\uparrow\downarrow }$} 
\newcommand{\uudduu}{Cr$^{\uparrow\uparrow}$/Cr$^{\downarrow\downarrow}$/Cr$^{
\uparrow\uparrow}$}
\newcommand{\uuuuuu}{Cr$^{\uparrow\uparrow}$/Cr$^{\uparrow\uparrow}$/Cr$^{
\uparrow\uparrow}$}

\begin{document}
\title{Magnetic and Electronic Switch in Metal Intercalated Two-Dimensional 
GeP$_3$} 

\author{D. P. de A. Deus,\textit{$^{a}$} I. S. S. de Oliveira,\textit{$^b$} 
J. B. Oliveira,\textit{$^c$} W. L. Scopel \textit{$^{d}$} and R. H. 
Miwa\textit{$^{e}$}}

\affiliation{$^a$Instituto Federal de Educa\c{c}\~ao, Ci\^encia e Tecnologia de 
Goi\'as, Departamento de \'Areas Acad\^emicas, Campus Jata\'i, 775 Orminda 
Vieira de Freitas, Jata\'i, GO, Brazil.}
\affiliation{$^b$Departamento de F\'isica, Universidade Federal de Lavras, 
	C.P. 3037, 37200-000, Lavras, MG, Brazil}
\affiliation{$^c$Instituto Federal do Tri\^angulo Mineiro, 38270-000, Ituiutaba, 
MG, 
Brazil}
\affiliation{$^d$Departamento de F\'{\i}sica, Universidade Federal do 
Esp\'irito 
Santo, Vit\'oria, ES, 29075-910, Brazil}
\affiliation{$^e$Instituto de F\'isica, Universidade Federal de Uberl\^andia,
        C.P. 593, 38400-902, Uberl\^andia, MG, Brazil}

\date{\today}
    
\begin{abstract}
Intercalation of foreign atoms in two dimensional hosts has 
been considered a quite promising route in order to engineer the electronic, 
and 
magnetic properties in 2D plataforms. In the present study, we performed a 
first-principles theoretical investigation of the energetic stability, and the  
magnetic/electronic properties of 2D \gep\, doped by Cr atoms. Our total energy 
results reveal the formation of   thermodynamically stable   Cr doped GeP$_3$ 
bilayer [\bl], characterized by interstitial Cr atoms lying in the  van der 
Waals (vdW) gap between \bl\, [\blcr]. We show that the ground state row-wise 
antiferromagnetic (RW-AFM) phase of \blcr\, can be tuned to a ferromagnetic  
(FM) configuration  upon  compressive mechanical strain ($\varepsilon$), 
\ud\,$\xrightarrow{\varepsilon}$\,\uu. By considering  its stacked 
counterparts, \blcr/\blcr\, and \blcr/Cr/\blcr, we found that  such a magnetic 
tuning is dictated by a combination of intralayer and interlayer couplings, 
where the RW-AFM phase change to layer-by-layer FM (\uuuu) and AFM (\uudduu) 
phases, respectively. Further electronic band structure calculations show that 
these Cr doped systems are metallic, characterized by the emergence  of strain 
induced  spin polarized channels at the Fermi level. These findings  reveal  
that the atomic intercalation, indeed, offers a new set of degree of freedom 
for the design and control the magnetic/electronic  properties in 2D 
systems.

\end{abstract}

\maketitle

\section{Introduction}

Many efforts have been devoted in the last years for the development of 
two-dimensional (2D) materials. The combination of quantum confinement and 
surface effects leads to novel physical properties, which are not present in 
their bulk counterparts. Since graphene was first isolated and 
characterized\,\cite{novoselovScience2004-2}, many others 2D materials with 
graphene-like structures have been synthesized or theoretically 
predicted\,\cite{C0NR00323A}, such as silicene, germanene, phosphorene, among 
others\,\cite{doi:10.1002/smll.201402041}, layered transition metal 
dichalcogenides\,\cite{doi:10.1021/cr300263a}, transition metal 
carbides\,\cite{doi:10.1002/adma.201805472}. These materials are envisioned to 
have a wide range of technological applications.  

Over recent years, it has been reported the structural stability of various 2D 
triphosphides. They all present interesting properties, with many possible 
technological applications. Miao et al. \cite{doi:10.1021/jacs.7b05133} 
demonstrated, through first principles simulations, the dynamical and chemical 
stability of monolayer InP$_3$, and they also verified ferromagnetic and 
half-metallic states induced by p-type doping. A second stable phase of InP$_3$ 
has also been recently proposed \cite{C9TC02030F}. BP$_3$ has also been predict 
to have a stable 2D layered phase \cite{C7TC02346D}, which is also the case for 
2D SnP$_3$ \cite{C8TA02494D,doi:10.1021/acs.jpcc.8b06668} and 2D CaP$_3$ 
\cite{doi:10.1021/acs.jpclett.8b00595}. In common, they all present moderate 
band gaps (adjustable through strain engineering), high carrier mobility  and 
strong light absorption in the visible and ultraviolet regions.  These features 
show promising applications as anode materials for Li-ion and Na-ion batteries 
\cite{C7TA10248H,doi:10.1021/acsaem.8b00621}, and also in electronic, 
spintronic, and photovoltaic devices. Moreover, Yao et al. \cite{YAO20195948} 
showed that SbP$_3$ and GaP$_3$ monolayers are also stable, and present larger 
band gaps than the other 2D triphosphides structures mentioned so far, and are 
suggested to be potential candidates for water-splitting photocatalysts. 

In particular, the bulk phase of germanium triphosphide (GeP$_3$) has long been 
synthesized \cite{DONOHUE1970143,GULLMAN1972441}, which consist of planes 
stacked through van der Waals interactions along the perpendicular direction. 
This material presents metallic behavior. Recently, Jing et al. 
\cite{doi:10.1021/acs.nanolett.6b05143} have shown that few layers GeP$_3$ are 
chemically, mechanically and dynamically stable, by means of first principles 
calculations. The monolayer and bilayer GeP$_3$ have a small indirect band gap 
(0.55 and 0.43\,eV respectively) while the trilayer becomes metallic. Moreover, 
the monolayer GeP$_3$ presents high carrier mobility and solar absorption. The 
combination of these remarkable properties with the prediction of 
experimentally 
accessible cleavage energies have motivated a large number of theoretical works 
proposing possible applications and new features of monolayer GeP$_3$. For 
example, Li et al. \cite{doi:10.1021/acsami.8b05655} have proposed another 
stable phase of monolayer GeP$_3$, and demonstrated that this system shows a 
ferromagnetic ground state through hole doping the material beyond a critical 
value. Monolayer GeP$_3$ has been shown to have high capacity, favorable ion 
hopping barrier, and robust structural stability, thus being a strong candidate 
for applications in electrodes of Li-ion \cite{C7CP04758D} and non-Li ion 
\cite{doi:10.1021/acs.jpcc.8b11574} batteries. Further investigations on strain 
engineering has also been recently carried out \cite{Zeng_2018}. In 
Refs.~\citenum{Wang2018} and \citenum{C8CP06310A} the electronic and transport 
properties of GeP$_3$ nanoribbons were investigated. GeP$_3$ monolayer may also 
find applications as gas sensors \cite{NIU201937,TIAN2019181}. In 
Ref.~\citenum{C9NR03255J} it is verified that GeP$_3$ monolayer is highly 
active 
for hydrogen evolution reaction, where the effect of a graphene substrate is 
also considered. Meanwhile, in a recent study, Zhang {\it et 
al.}\,\cite{zhang2019strong} performed a detailed study of  the electronic 
properties, and  the energetic stability of \gep\, monolayer doped by elemental 
impurities.

To broaden the application possibilities of 2D crystals there is nowadays an 
increasing interest to obtain 2D materials with magnetic 
properties\cite{gong2019two}. Recently, two chromium compounds have been 
verified to be ferromagnetic: Cr$_2$Ge$_2$Te$_6$ \cite{Gong2017} and CrI$_3$ 
\cite{Huang2017}. Currently, many others 2D magnetic crystals have been 
obtained 
\cite{Gongeaav4450,Gibertini2019}. Control of the magnetic properties in 2D 
systems has been pursued based on the proximity effects\cite{Gong2017}. For 
instance, in vdW stacking of CrI$_3$ bilayers\cite{jiang2018electric} where the 
magnetic coupling between the stacked layers depends on the CrI$_3$/CrI$_3$ 
interface geometry\cite{sivadas2018stacking,jiang2019stacking}. Further studies 
examined the tunning of the magnetic phases in CrI$_3$ monolayer and bilayers 
as 
a function of an external electric field 
\cite{jiang2018electric,morell2019control}.

In the present study, we show that magnetic and electronic properties of few 
layers of \gep\, doped by Cr atoms can tuned by mechanical strain. Based on 
first-principles calculations,  we investigate the energetic stability, 
magnetic, and electronic  properties of the \gep\ bilayer  [\bl] doped with 
interstitial Cr atoms [\blcr], and its stacked counterparts, namely 
\blcr/\blcr, 
and the Cr intercalated \blcr/Cr/\blcr\ structure. We show that these systems 
present  tuneable  magnetic phases, characterized by an intralayer AFM to FM 
transition and  interlayer FM and AFM couplings, respectively.
Further electronic band structure calculations reveal the emergence of strain 
induced spin polarized channels at the Fermi level, suggesting that \blcr\, is 
an interesting building block to the design 2D spintronic devices.

\section{Methods}

The calculations were performed by using the Density Functional Theory (DFT) as 
in implemented in the {\sc Quantum ESPRESSO}\,\cite{espresso} code. The 
exchange 
correlation energy was described  within the spin-polarized generalized 
gradient 
approximation, as proposed by Perdew, Burke and Ernzerhof 
(GGA-PPE)\,\cite{pbe}. 
The electron-ion interactions were described using the projector 
augmented wave potential (PAW)\,\cite{paw}. The Kohn-Sham wave functions were 
expanded in a plane-wave basis set with an energy cutoff of 48 Ry for the 
single 
particle wave functions, and  457 Ry for the total charge density. The atomic 
positions and lattice vectors  were fully optimized by including van der Waals 
(vdW) interactions by using  the vdW-DF 
approach\,\cite{thonhauserPRL2015,thonhauserPRB2007,berlandRepProgPhys2015, 
langrethJPhysC2009}, where we have considered a force 
convergence criterion of  1\,mRy/bohr 
and 10$^{-5}$~eV for the total energy.   We have checked the convergence of our 
results by  increasing the energy cutoffs to 60\,Ry (plane basis-set) and 
480\,Ry (total charge density), and reducing the force convergence criterion to 
0.2\,mRy/bohr. The few layers \gep\, systems were described within the 
supercell 
approach, where we have employed a vacuum region of (at least) 20~\AA\  between 
neighboring slabs, that is, normal to the \gep\, layers. The Brillouin zone  
sampling was performed by using a 3$\times$3$\times$1  Monkhorst-Pack \cite{mp} 
mesh. Due to the nature of d orbitals, we have used the GGA+U method for 
simulating the doped structures; where we set an effective interaction 
parameter $U_{\rm eff}=2$\,eV of Cr(3d) electrons. In addition, some key 
results were further confirmed by using the Vienna 
Ab initio Simulation Package (VASP).

\section{Results and Discussions}

\subsection{\label{bilayer} Pristine  \bl}

Firstly, as a benchmark, we compare our results of equilibrium geometry and 
total energies of \bl\, with the ones presented in the seminal work on 2D 
\gep\, 
performed by Jing {\it et al.}\,\cite{jinggep3}. The pristine \gep\, bilayer 
presents  a puckered hexagonal structure, in each layer the Ge atoms bonds to 
three neighboring P atoms, and each P atom forms two P--P bond and one Ge--P 
bond,  \Figs{model1}(a) and (b). At the equilibrium geometry, we 
found optimized lattice constants of $a=b=7.08$\,\AA,  P--P and Ge--P bond 
lengths of 2.25 and 2.50\,\AA, respectively.  The  \gep\, MLs are 
separated by $d=2.09$\,\AA, and the vertical distance between the topmost and 
bottommost Ge atoms ($h$) is equal to 5.05\,\AA, as shown in 
Fig.\ref{model1}(a). In order to verify the energetic stability of the bilayer 
system, we compare the total energies of \gep\, BL  ($E_{\rm BL}$) and  ML 
($E_{\rm ML}$), $\Delta E_{\rm BL}= E_{\rm ML}-E_{\rm BL}/2$. We found that the 
formation of \bl\, is an exothermic process, $\Delta E_{\rm 
BL}$\,=\,52~meV/\AA$^2$. These 
findings are in good agreement with the ones in Refs.\cite{jinggep3,KIM2018126}.

\subsection{\label{bilayer_Cr} Cr-doped GeP$_3$ bilayer}

The electronic and magnetic properties in 2D systems have been engineered upon 
the atomic intercalation processes\,\cite{wang2015physical}. For instance, by 
the  incorporation of foreign atoms\,\cite{gong2018spatially}, or  taking 
advantage  of the symmetry resulting from the stacked layers, giving rise to 2D 
atomic self assemblies\,\cite{miwa2015periodic,zhao2020engineering}. Here, we 
examine the energetic stability, the magnetic/electronic  properties of 
\bl\, doped by Cr atoms, and  the control of  these properties by external 
pressure.

\begin{figure}[!htp]
\centering
 \includegraphics[width=\columnwidth]{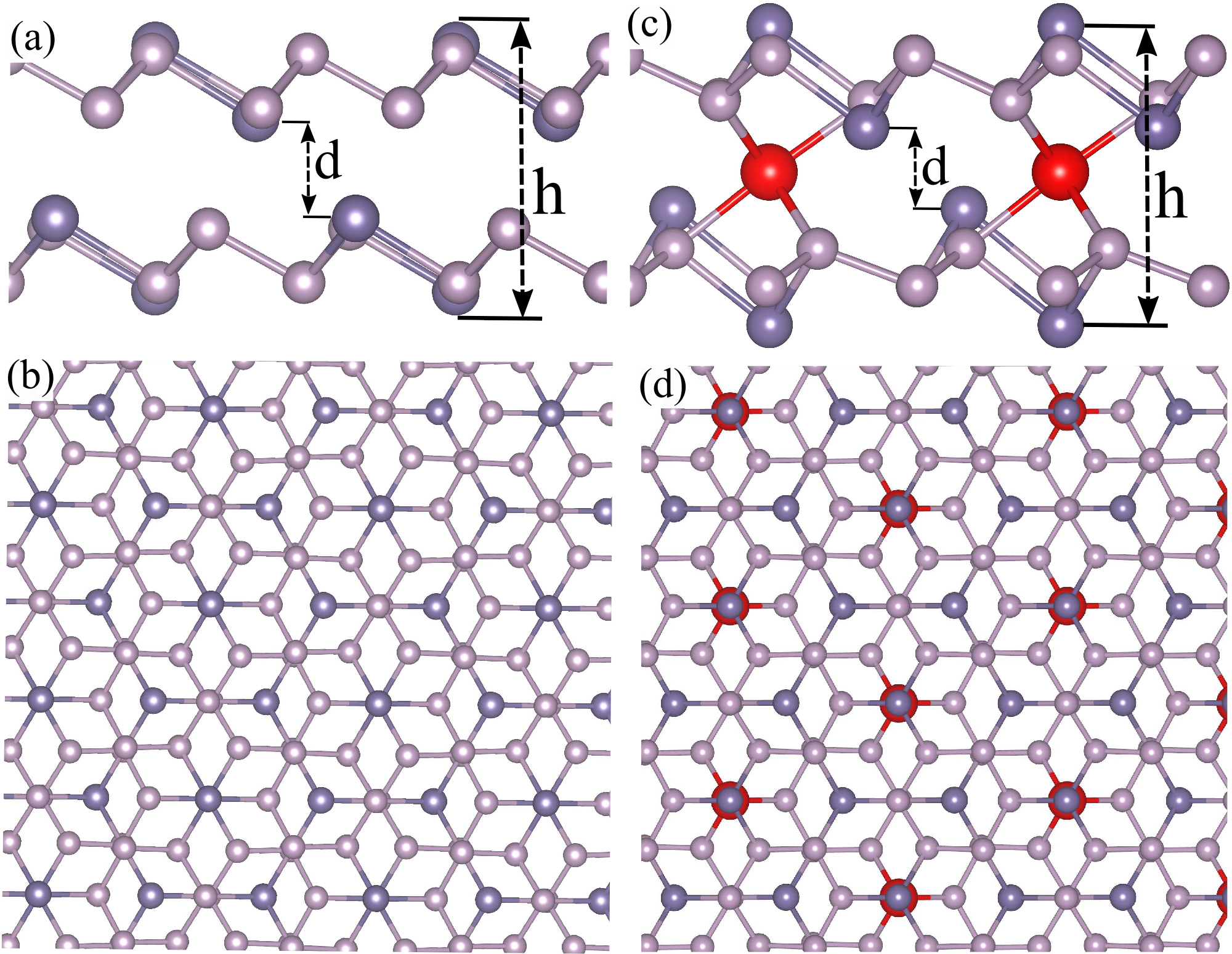}
   \caption{Structural model of pristine (left)  and Cr doped (right) \bl; side 
 [top]  views (a) and (c) [(b) and (d)]. The blue, gray 
and red color indicates the germanium, phosphorus and chromium atoms, 
respectively. The separation between top and bottom Ge atoms on \bl\ and 
\blcr, as well as the spacing between the individual layers in GeP$_{3}$, are 
indicated by $h$ and $d$, respectively.}
   \label{model1}
 \end{figure}

Firstly, we have considered a number of plausible configurations for Cr atoms 
in 
\bl. In Fig.\,\ref{model1}(c) we present the lowest energy configuration, where 
we find the Cr atoms occupying  the rhombohedral room between the GeP$_3$ MLs 
[\blcr].  The energetic stability of such a structural model was examined  
through the calculation of  the formation energy ($E^f[{\rm X}]$),
\begin{equation}
E^f[{\rm X}]=\frac{1}{n_{\rm ML}}(E_{\rm X} - n_{\rm ML}\times E_{\rm ML} 
-n_{\rm Cr}\times\mu_{\rm Cr}),
\end{equation}
where $E_{\rm X}$ is the total energy of the final system, here X=\blcr, and 
$E_{\rm ML}$ is the total energy of the energetically stable  \gep\, 
ML\,\cite{jinggep3}. $n_{\rm ML}$ and $n_{\rm Cr}$ are the number of \gep\, MLs 
and the Cr atoms within our supercell approach, and $\mu_{\rm Cr}$ is the 
chemical potential of Cr atoms. The upper limit of $\mu_{\rm Cr}$ is the 
chemical potential of its elemental (bcc bulk) phase, $\mu_{\rm Cr} \leq 
\mu_{\rm Cr}^{\rm bulk}$. By considering the  upper limit of $\mu_{\rm Cr}$, 
\,$\mu_{\rm Cr}=\mu_{\rm Cr}^{\rm bulk}$ (Cr rich condition), we found that 
\blcr\, is an energetically stable structure with  $E^f=-20$\,meV/\AA$^2$. 
Further structural stability  was confirmed by performing {\it ab initio} 
molecular dynamics (AIMD) simulations at 300\,K. We have considered time steps 
of 5\,fs, and we observe that the \blcr\, structure has been preserved up to 
10\,ps. In Fig.\,\ref{aimd} (appendix) we present the total energy fluctuation  
as a function of the simulation time. 

The interstitial Cr atoms  embedded between the \gep\, layers form a triangular 
lattice [Fig.\,\ref{model1}(d)], giving rise to a $\delta$-doping like 
structure.  We found that the lattice constants, $a=b=7.05$\,\AA, and the 
interlayer distance, $d$=2.11\,\AA, of \blcr\,  are practically the same 
compared with those of pristine \bl, $a=b=7.08$\,\AA, and $d=2.09$\,\AA, thus 
indicating that the local atomic structure of \bl\ is weakly  perturbed by the 
presence of 
interstitial Cr atom. At the equilibrium geometry, we found Cr--P and Cr--Ge 
bond lengths of 2.59 and 3.17\,\AA, respectively, which are slightly larger than 
the sum of 
their covalent radii, increasing the vertical distance $h$  from 5.05\, to 
6.34\,\AA, Figs.\,\ref{model1}(a) and (c).

\begin{figure}[!htp]
\centering
  \includegraphics[width=\columnwidth]{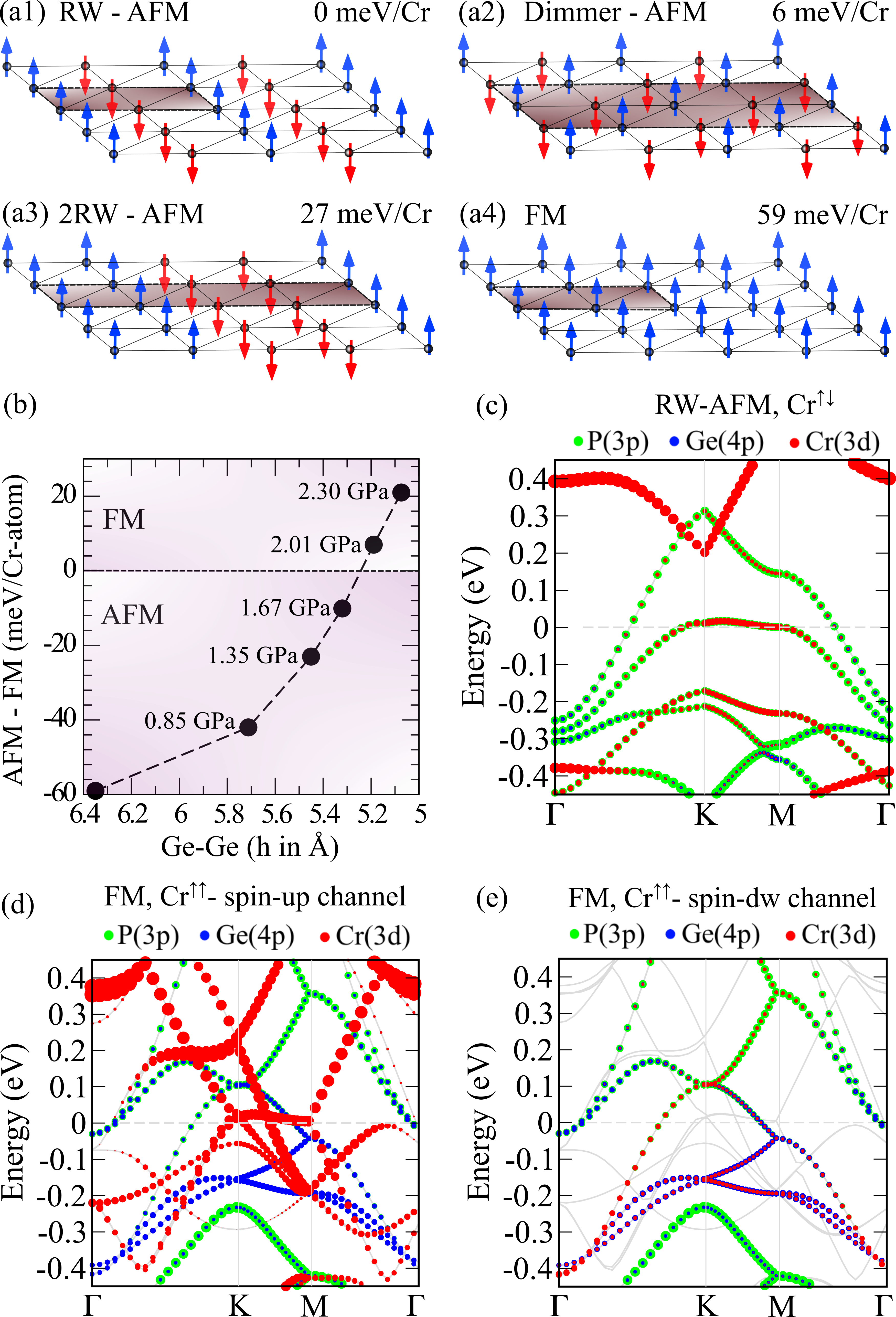}
   \caption{Magnetic configurations models of \blcr, (a1) row-wise 
antiferromagnetic (RW-AFM), (a2) Dimmer-AFM,  (a3) 2RW-AFM, and (a4) 
ferromagnetic (FM) phases. (b) Total energy difference between RW-AFM and FM 
phases as a function of the external compressive strain. Orbital 
projected electronic band structures of the uncompressed ($h=6.34$\,\AA) 
RW-AFM \blcr\,  (c),  and compressed ($h=5.07$\,\AA) FM \blcr, spin-up (d) and 
-down (e) channels.}
   \label{spin-conf}
 \end{figure}
 
The interstitial Cr atoms present a net magnetic moment of  
$3.4\,\mu_{\rm B}$, which are coupled  mediated by the P atoms. We have 
compared  the energetic stability of four magnetic 
phases\,\cite{qiuJPhysChemC2019}, as shown in Figs.\,\ref{spin-conf}(a1)--(a4). 
At the equilibrium geometry, we found the row-wise antiferromagnetic (RW-AFM) 
phase [Fig.\,\ref{spin-conf}(a1), hereafter labeled as \ud] as the most stable 
spin-configuration, and   the ferromagnetic  (FM) phase 
[Fig.\,\ref{spin-conf}(a4), hereafter labeled as \uu] as the least stable one 
by $E_{\text{RW-AFM}}-E_{\rm FM}=-59$\,meV/Cr-atom. On the other hand, we found 
that such a energetic preference for the RW-AFM alignment in \blcr\, can be 
tuned by mechanical compressive strain ($\varepsilon$). In 
Fig.\,\ref{spin-conf}(b) we present our results of $E_{\text{RW-AFM}}-E_{\rm 
FM}$ as a function of the vertical distance $h$.  It is noticeable a 
\ud\,$\xrightarrow{\varepsilon}$\,\uu\, magnetic transition for $h$ between 
5.32 and 5.19\,\AA, and upon further compression to $h=5.07$\,\AA, which 
corresponds to an external pressure of 2.30\,GPa, the FM phase become more 
stable by $E_{\text{RW-AFM}}-E_{\rm FM}=+21$\,meV/Cr.

The orbital projected electronic band structure of the RW-AFM \blcr, 
Fig.\,\ref{spin-conf}(c), reveals the formation of metallic  bands mostly ruled 
by the P(3p), while the Cr atoms give rise to localized  states  resonant 
with the Fermi level along the KM direction. In contrast, the spin-up and -down 
energy bands  are no longer degenerated upon compression, 
Figs.\,\ref{spin-conf}(d) and (e). The  spin-up channel presents a larger 
density 
of states  near the Fermi level compared with that of the spin-down channel. 
The 
former can be characterized by Cr(3d)  bands  along the  KM direction, one 
dispersionless and the other with nearly linear dispersion crossing the Fermi 
level. Meanwhile,   both spin-channels present nearly parabolic metallic bands  
 
along the $\Gamma$M and $\Gamma$K directions, which are projected on the Ge(4p) 
and P(3p) orbitals. These findings allow us to infer  the emergence of 
mechanically tuneable  spin polarized electronic transport along the \blcr\, 
layers dictated by the spin-polarized Cr(3d) states near the Fermi level.

\subsection{${\bf (GeP_3)_{BL}^{Cr}}$~interfaces}

Stacking of 2D magnetic system is a quite interesting option aiming the  search 
 
of new  phenomena ruled by interlayer interactions. For 
instance, magnetic phases in few layers  CrI$_3$ as a function of the stacking  
geometry\,\cite{jiang2019stacking}, that in turn  can be tuned by  external 
agents like electric field\,\cite{morell2019control} or mechanical 
pressure\,\cite{li2019pressure,song2019switching}. Here, we investigate the 
tuning of the magnetic properties of \blcr\,  bilayers mediated by vertical 
compressive strain.  Firstly we have considered a pristine bilayer system, 
\blcr/\blcr, and then the intercalation of  Cr atoms, \blcr/Cr/\blcr. It is 
worth noting that such a compression will promote two  concurrent magnetic 
interactions, (i) the intralayer RW-AFM\,$\rightarrow$\,FM transition discussed 
above; and (ii) proximity effects at  the \blcr--\blcr\, 
interface.
\begin{figure}[ht]
\centering
  \includegraphics[width=\columnwidth]{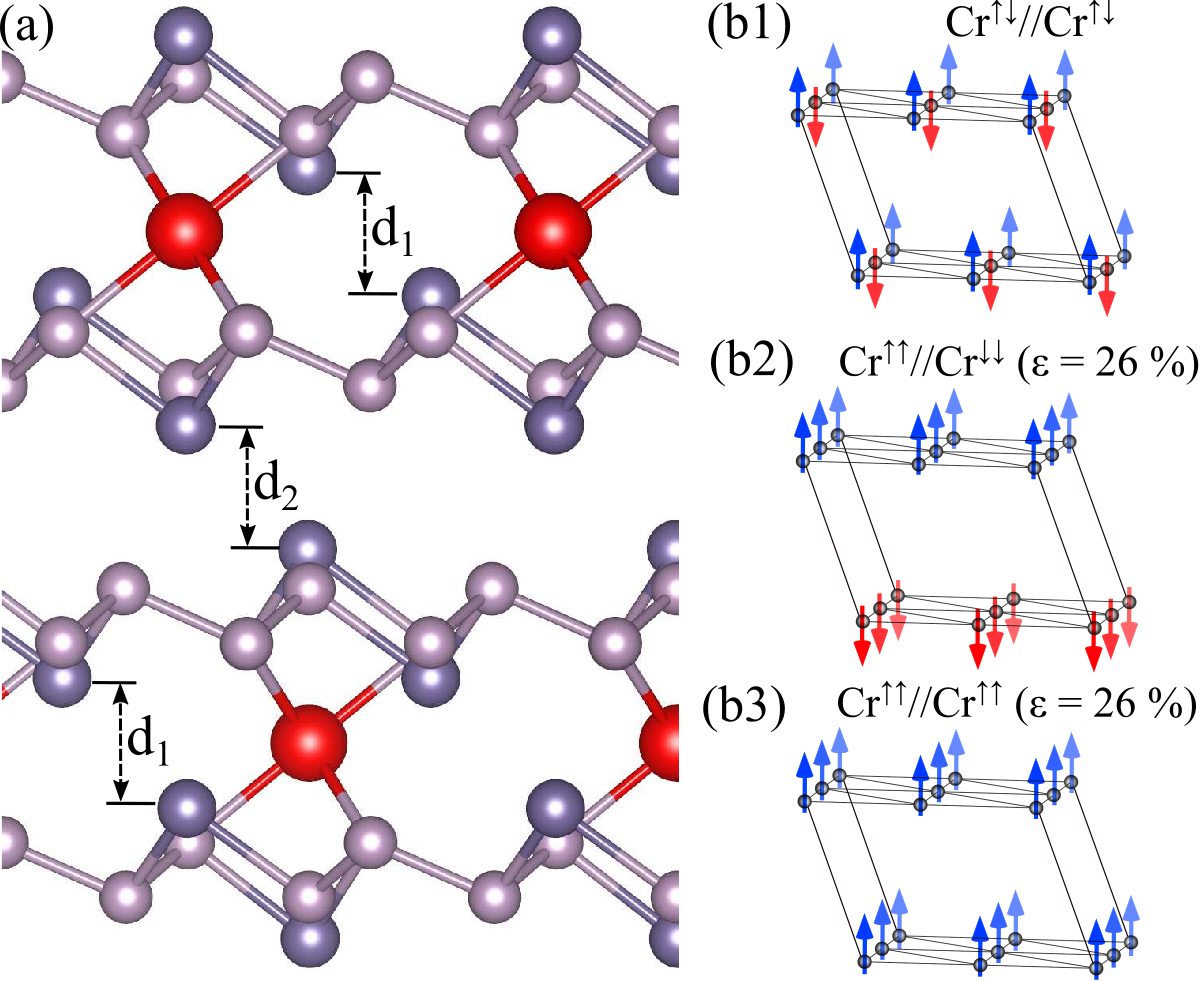}
  \caption{(a) Structural model of \blcr/\blcr, and the schematic spin 
configurations  \udud\, (b1), \uudd\, (b2), and \uuuu\, (b3). The localization 
and the spin-polarization of the Cr-atom are indicated by arrows.  The 
interface distances are indicated by d$_1$ and d$_2$ in (a).}
  \label{fig:model2Cr}
\end{figure}

In \Fig{fig:model2Cr}(a) we present the structural model of \blcr/\blcr, where  
the \gep\, layers are stacked following the  conventional ABC stacking order  
of 
its bulk phase\,\cite{doi:10.1021/acs.nanolett.6b05143}. At the equilibrium 
geometry, the structural properties of each \blcr\ are mostly preserved, for 
instance the vertical distances $d_1$ and $h$ of 2.06  and 6.44\,\AA, 
respectively. The  interlayer spacing between the two \blcr\, layers, 
$d_2=2.07$\,\AA, also is practically the same compared with that of pristine 
\bl. There are no chemical bonds connecting the \blcr\, layers, indicating that 
the energetic stability of the stacked system is ruled by  vdW interactions. 
The 
energetic stability of the bilayer system was verified by the 
calculation of the formation energy. For X=\blcr/\blcr\, in eq.\,(1), we 
obtained $E^f=-25$\,meV/\AA$^2$.

Focusing on the magnetic properties,  we have examined three plausible magnetic 
configurations, namely, (i)  each  \blcr\, layer presents RW-AFM 
alignment (\udud), (ii) the \blcr\, layers present intralayer FM 
coupling, and  interlayer AFM coupling (\uudd), and (iii) the \blcr\, layers 
present intralayer and interlayer  FM coupling (\uuuu). These spin 
configurations are schematically shown in \Figs{fig:model2Cr}(b1)--(b3).  Our 
total energy results reveal  the former configuration, \udud, 
as the most stable one, followed by  \uudd\, and \uuuu\, by 51 and 67\,meV/Cr, 
respectively. The energy difference of 16\,meV between these two intralayer FM  
phases indicates the presence of magnetic coupling between the FM \blcr\, 
layers,  favoring the  former one. 
%%%%%%
\begin{table}[h]
  \centering
  \caption{\label{enertbl} Total energy differences (in meV/Cr-atom) with 
respect to the energetically more stable spin configuration of the free ($\rm 
\Delta E_{Free}$) and compressed ($\rm\Delta E_{Compress}$)  \blcr//\blcr\, 
and \blcr/Cr/\blcr\, systems.}
\begin{ruledtabular}
  \begin{tabular}{ccc}
%     \hline\hline
structure & $\rm \Delta E_{Free}$ & $\rm \Delta E _{Compress}$  \\
\hline
     \udud\,  &      0            &    4    \\
     \uudd\,  &     51            &    39    \\
     \uuuu\,  &     67            &    0     \\
\hline
     \ududud\, &    0             &   38  \\
     \uudduu\, &   60             &   0   \\
     \uuuuuu\, &   20             &   33  \\
     %    \hline\hline
  \end{tabular}
\end{ruledtabular}
\end{table}
%%%%%%5

 \begin{figure*}
 \centering
 \includegraphics[width=2\columnwidth]{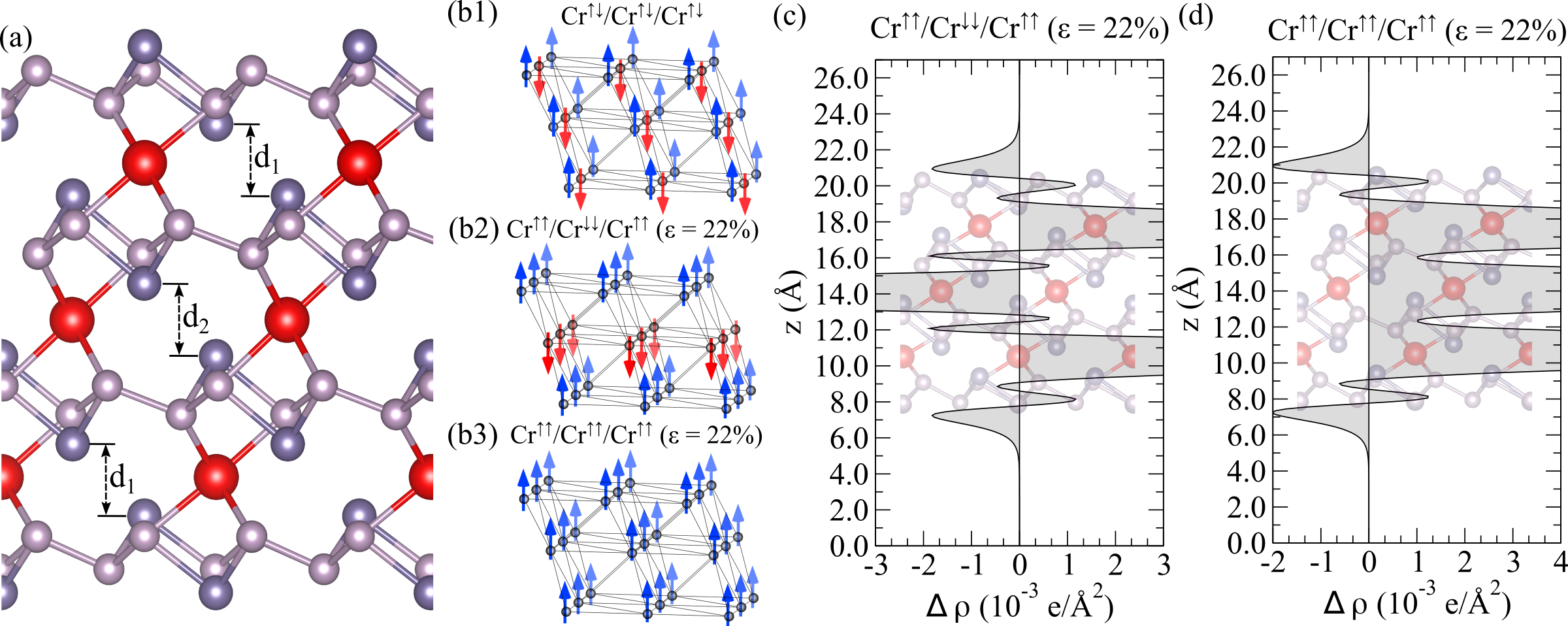}
 \caption{(a) Structural model of \blcr/Cr/\blcr, and  the spin 
configurations \ududud\, (b1),  \uudduu\, (b2), and \uuuuuu\, (b3). The 
localization and the spin-polarization of the Cr-atom are indicated by arrows. 
Planar averaged spin-density along the $z$-axis of the compressed 
\blcr/Cr/\blcr\, system, \uudduu\, (c) and \uuuuuu\, (d). The 
interface distances are indicated by d$_1$ and d$_2$ in (a).} 
 \label{fig:estru_strain_magn}
 \end{figure*}

Such a magnetic coupling can be strengthened by a compressive strain normal to 
the stacking direction. In this case, concomitantly with the reduction of the 
vertical interlayer distance d$_2$, the reduction of d$_1$  will favor the 
intralayer FM interaction between the Cr atoms, 
\ud\,$\xrightarrow{\varepsilon}$\,\uu. Indeed, for a compressive strain of 
$\varepsilon$ = 26\,\%, which corresponds to a external pressure ($P$) of 3.82 
GPa,  we found  energetic preference for the intralayer and interlayer FM 
configuration [Fig.\,\ref{fig:model2Cr}(b3)];  thus characterizing a 
\udud$\xrightarrow{\varepsilon}$\,\uuuu\, magnetic transition tuned  by a  a 
mechanical  strain. Our total energy results are summarized in Table\,I.  These 
results, in addition to the spin polarization of the Ge and P atoms at the 
interface region, suggest an indirect magnetic interaction between the \blcr\, 
layers. 

Such a magnetic coupling between the  \blcr\, layers can be 
strengthened by the intercalation  of Cr atoms, \blcr/Cr/\blcr. The Cr-atoms 
lie 
in the rhombohedral room between the \blcr\, layers 
[\Fig{fig:estru_strain_magn}(a)], that is,  the  same geometry as obtained for 
the isolated \blcr. By using the eq.(1), for X=\blcr/Cr/\blcr, we found 
$E^f$ of $-21$\,meV/\AA$^2$, thus supporting the energetic 
stability of the Cr $\delta$-doped GeP$_3$ quadrilayers [\qlcr]. At the 
optimized  geometry, \qlcr\,  presents interlayer distances of $d_1=2.00$~\AA, 
and $d_2=2.04$~\AA, while the undoped pristine GeP$_3$ quadrilayer presents 
equally spaced interlayer distances of 1.97\,\AA. 

Next we investigate the magnetic couplings between the Cr atoms. We have 
considered the spin configurations presented in 
\Figs{fig:estru_strain_magn}(b1)--(b3), hereafter labeled as \ududud, \uudduu\, 
and \uuuuuu. At the equilibrium geometry, we found the former spin 
configuration, where each \blcr\, is characterized by the RW-AFM alignment, 
more 
stable than the other ones by 20 and 60\,meV/Cr-atom, respectively. However, 
for 
$\varepsilon$ of 22\,\% ($P$ = 4.09\,GPa),  \uudduu\, 
[Fig.\ref{fig:estru_strain_magn}(b2)] becomes more stable than \uuuuuu\, and 
\ududud\, by 33 and 38\,meV/Cr-atom, as presented in Table\,I. Thus, revealing 
that \blcr/Cr/\blcr\, also presents a tuneable spin configuration, 
\ududud$\xrightarrow{\varepsilon}$\,\uudduu. However, differently from its 
\blcr/\blcr\, counterpart, the spin-polarizations are alternated along the 
stacked layers. 

In order to provide a more complete picture of the interlayer   coupling in 
\qlcr,  we have examined  the planar averaged  spin-density 
[$\Delta\rho^{\uparrow\downarrow}(z)$] along the stacking direction ($z$), 
\begin{equation} \Delta\rho^{\uparrow\downarrow}(z) = \frac{1}{S}\int_{S} 
[\rho^\uparrow(x,y,z) - \rho^\downarrow(x,y,z)]dxdy, \end{equation} where $S$ 
represents the surface area in the $xy$-plane, and 
$\rho^\uparrow/\rho^\downarrow$ are the spin-polarized total charge densities. 
Our results of $\Delta\rho^{\uparrow\downarrow}(z)$ for the spin configurations 
\uudduu, and 
\uuuuuu\, at $\varepsilon$ = 22\% are shown in 
Figs.\,\ref{fig:estru_strain_magn}(c) and (d). The  net magnetic moments are 
mostly localized in the layers containing Cr atoms, however in the \uudduu\, 
configuration, the interface P atoms also become spin polarized. The energetic 
advantage of \uudduu, in the compressed system, comes from a combination of  
(i) 
the \ud\,$\xrightarrow{\varepsilon}$\,\uu\, transition within the \blcr\, 
units, 
and (ii) the lowering of the kinetic energy ruled by a super-exchange 
interaction, mediated by the interface P  atoms 
[Figs.\,\ref{fig:estru_strain_magn}(c)]. As shown in 
Figs.\,\ref{fig:estru_strain_magn}(d), such (kinetic) energy gain is not 
allowed in \uuuuuu.

\begin{figure}
\centering
  \includegraphics[width=\columnwidth]{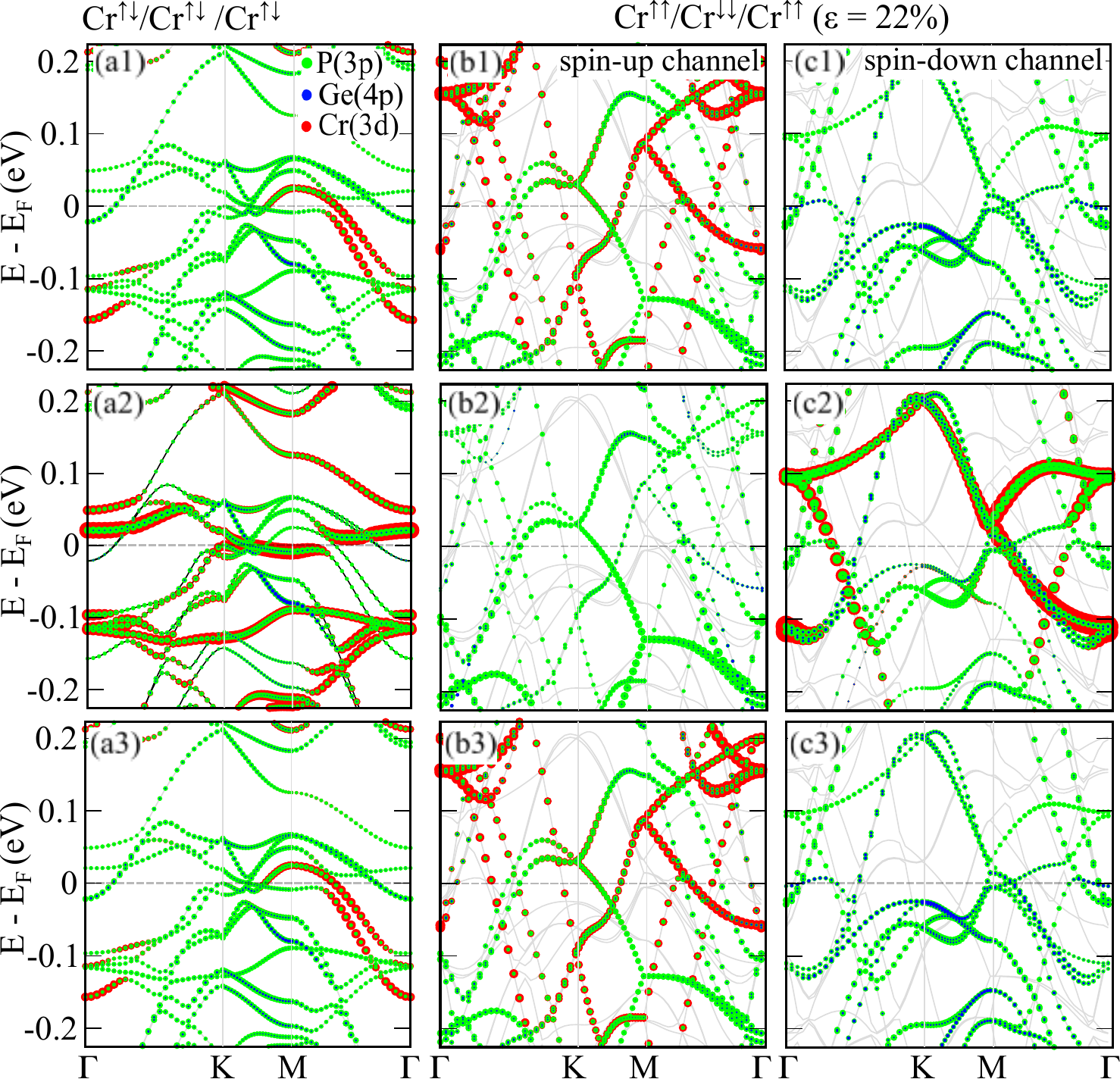}
  \caption{The orbital projected electronic band 
structures of  uncompressed (a) \ududud\,   and  compressed [(b) spin-up, 
and (c) spin-down] \uudduu\, systems, projected on the top (a1)-(c1), central 
(a2)-(c2), and bottom (a3)-(c3) layers of  \blcr/Cr/\blcr.}
  \label{fig:banda-4ml-3Cr}
\end{figure}

Finally, focusing on the electronic properties, in 
Fig.\,\ref{fig:banda-4ml-3Cr}(a) we present the orbital projected electronic 
band structure of the uncompressed \qlcr. The projection on the edge \blcr\, 
layers present the  same band structures [Figs.\,\ref{fig:banda-4ml-3Cr}(a1) 
and 
(a3)], where we find the formation of metallic bands, mostly composed by Cr(3d) 
and P(3p) hybridized states. Meanwhile, the electronic bands  projected in the 
central layer are  characterized by  the strengthening  of dispersionless 
Cr(3d) 
bands near the Fermi level, Fig.\,\ref{fig:banda-4ml-3Cr}(a2). Since we have 
considered the lowest energy configuration, \ududud, of the uncompressed system 
($\varepsilon$\,=\,0), the  spin-up and -down channels are degenerated. In 
contrast, such a degeneracy has been removed for $\varepsilon=22\,\%$,  where 
the \uudduu\, configuration becomes more stable.  As shown in   
Figs.\,\ref{fig:banda-4ml-3Cr}(b1) and (b3), the hybridization of Cr(3d) and 
P(3p) orbitals gives rise to metallic bands along the edge \blcr\, layers. In 
contrast, the density of metallic states has been reduced in the central layer, 
with no projection of the Cr(3d) orbitals, Fig.\,\ref{fig:banda-4ml-3Cr}(b2). 
The spin-down energy bands present a somewhat  opposite picture, namely the  
Cr(3d) and P(3p) hybridized bands  localized in the central \blcr\, layer 
[Fig.\,\ref{fig:banda-4ml-3Cr}(c2)], and no Cr(3d) orbital projection  on the 
spin-down channels at  the edge layers [Figs.\,\ref{fig:banda-4ml-3Cr}(c1) and 
(c3)]. These findings reveal that, in addition to the  spin configurations, the 
rise and (layer) localization of the spin-polarized metallic bands can also be 
tuned by  mechanical strain in Cr intercalated \gep\, systems.

\section{Conclusions}
We have performed an {\it ab initio} investigation of few layer systems of 
\gep\, doped by Cr atoms. The energetic and structural stabilities of Cr  atoms 
lying in the rhombohedral room  of between  \gep\, layers, \blcr, have been 
verified through  formation energy calculations and first-principles molecular 
dynamic simulations. We show that  \blcr, as well as its stacked counterparts 
are quite interesting platforms for tuneable magnetism  in 2D systems. We found 
that the ground state row wise antiferromagnetic configuration in \blcr, \ud,  
can be tuned to a ferromagnetic phase mediated by compressive mechanical strain 
($\varepsilon$), \ud\,$\xrightarrow{\varepsilon}$\,\uu. Meanwhile, in the 
stacked systems,  \blcr/\blcr, and \blcr/Cr/\blcr, the  combination of 
intralayer and interlayer interactions gives rise to the following strain 
induced changes in the magnetic configurations, 
\udud\,$\xrightarrow{\varepsilon}$\,\uuuu\, and 
\ududud\,$\xrightarrow{\varepsilon}$\,\uudduu, respectively. Concomitantly with 
such a magnetic tuning, further electronic structure calculations revealed  the 
emergence of spin polarized channels near the Fermi level. These findings 
indicate   that mechanical tuning of the  electronic and magnetic properties,  
in two dimensional platforms, can be engineered throughout synergic  
combinations between the foreign atoms and the 2D-hosts.

\section{Appendix}

\begin{figure}[htpb]
\centering
\includegraphics[width=\linewidth]{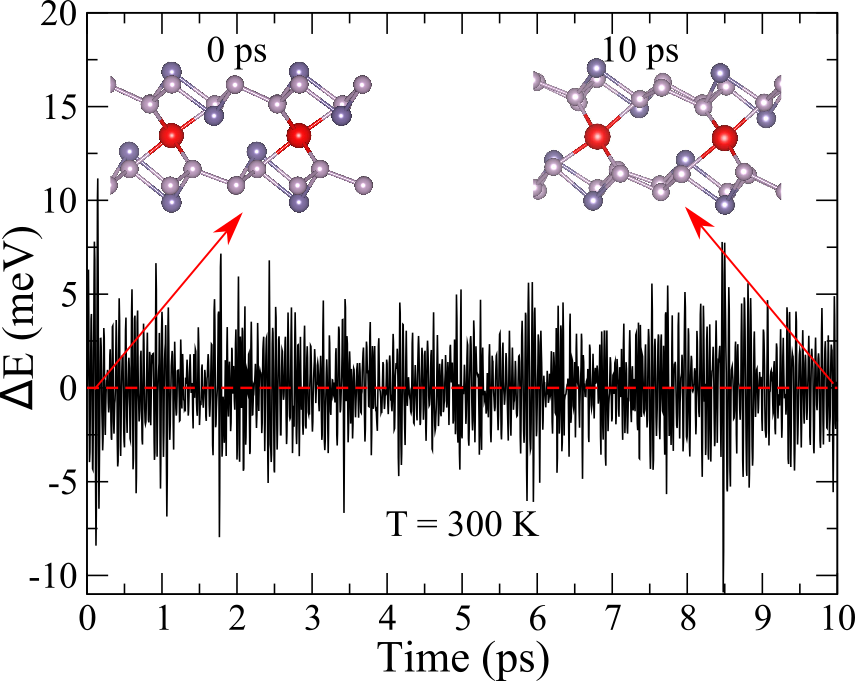} \caption{The total energy 
fluctuations as a function of time step (5\,fs) for the  AIMD simulation at 
 300\,K, of the RW-AFM \blcr.   The inset figures represent snapshots of atomic 
structures of \blcr\,  at 0\,ps (initial configuration) and  10\,ps (final 
configuration).} 
\label{aimd} 
\end{figure}

In order to provide further support to the energetic stability of the Cr 
intercalated \bl\, structure,  we performed ab-initio molecular dynamics (AIMD) 
simulations of the  \blcr\, system, in order to verify its thermal stability at 
room temperature. We have considered the ground state row-wise 
antiferromagnetic 
RW-AFM phase by using a 2$\times1$ surface periodicity, containing two formula 
units. The AIMD simulations were carried out using Nos\'e thermostat method at 
a 
temperature of 300 K for a total of 10 ps, with a time step of 5 fs. There is no 
 
bond breaking between Ge and P atoms during the AIMD simulation, after heating 
the
\blcr\, for 10~ps. Likewise, the individual layers of GeP$_3$ remained 
coupled via the Cr atoms during the simulation, as can be seen in the initial 
and 
final configuration of the structure [inset of Fig.\,\ref{aimd}], thus ensuring 
that it  is stable at room temperature.  The calculated total magnetic moment 
per (GeP$_3$)$^{Cr}_{BL}$ supercell varied between $\pm$0.22 
$\mu_{b}$/unit-cell 
during the AIMD simulation, showing that the RW-AFM spin configuration also 
keeps practically unchanged during the MD process.  From an energetic point of 
view, the total energy at 0\,ps and after 10\,ps has changed from -11.3 to 
10.8\,meV/unit-cell, Fig.\,\ref{aimd}, showing that the structural properties 
are maintained even at 300 K, enabling future experimental applications.

\begin{acknowledgments}

The authors acknowledge financial support from the Brazilian agencies CNPq, CAPES, FAPES and FAPEMIG and the CENAPAD-SP
%and LNCC (Laboratório Nacional de Computação ̃Científica %Projeto SCAFMat2) for computer time.

\end{acknowledgments}

\bibliography{gep3-cr_1}

\providecommand{\noopsort}[1]{}\providecommand{\singleletter}[1]{#1}%
\begin{thebibliography}{53}%
\makeatletter
\providecommand \@ifxundefined [1]{%
 \@ifx{#1\undefined}
}%
\providecommand \@ifnum [1]{%
 \ifnum #1\expandafter \@firstoftwo
 \else \expandafter \@secondoftwo
 \fi
}%
\providecommand \@ifx [1]{%
 \ifx #1\expandafter \@firstoftwo
 \else \expandafter \@secondoftwo
 \fi
}%
\providecommand \natexlab [1]{#1}%
\providecommand \enquote  [1]{``#1''}%
\providecommand \bibnamefont  [1]{#1}%
\providecommand \bibfnamefont [1]{#1}%
\providecommand \citenamefont [1]{#1}%
\providecommand \href@noop [0]{\@secondoftwo}%
\providecommand \href [0]{\begingroup \@sanitize@url \@href}%
\providecommand \@href[1]{\@@startlink{#1}\@@href}%
\providecommand \@@href[1]{\endgroup#1\@@endlink}%
\providecommand \@sanitize@url [0]{\catcode `\\12\catcode `\$12\catcode
  `\&12\catcode `\#12\catcode `\^12\catcode `\_12\catcode `\%12\relax}%
\providecommand \@@startlink[1]{}%
\providecommand \@@endlink[0]{}%
\providecommand \url  [0]{\begingroup\@sanitize@url \@url }%
\providecommand \@url [1]{\endgroup\@href {#1}{\urlprefix }}%
\providecommand \urlprefix  [0]{URL }%
\providecommand \Eprint [0]{\href }%
\providecommand \doibase [0]{http://dx.doi.org/}%
\providecommand \selectlanguage [0]{\@gobble}%
\providecommand \bibinfo  [0]{\@secondoftwo}%
\providecommand \bibfield  [0]{\@secondoftwo}%
\providecommand \translation [1]{[#1]}%
\providecommand \BibitemOpen [0]{}%
\providecommand \bibitemStop [0]{}%
\providecommand \bibitemNoStop [0]{.\EOS\space}%
\providecommand \EOS [0]{\spacefactor3000\relax}%
\providecommand \BibitemShut  [1]{\csname bibitem#1\endcsname}%
\let\auto@bib@innerbib\@empty
%</preamble>
\bibitem [{\citenamefont {Novoselov}\ \emph {et~al.}(2004)\citenamefont
  {Novoselov}, \citenamefont {Geim}, \citenamefont {Morozov}, \citenamefont
  {Jiang}, \citenamefont {Zhang}, \citenamefont {Dubonos}, \citenamefont
  {Grigorieva},\ and\ \citenamefont {Firsov}}]{novoselovScience2004-2}%
  \BibitemOpen
  \bibfield  {author} {\bibinfo {author} {\bibfnamefont {K.~S.}\ \bibnamefont
  {Novoselov}}, \bibinfo {author} {\bibfnamefont {A.~K.}\ \bibnamefont {Geim}},
  \bibinfo {author} {\bibfnamefont {S.~V.}\ \bibnamefont {Morozov}}, \bibinfo
  {author} {\bibfnamefont {D.}~\bibnamefont {Jiang}}, \bibinfo {author}
  {\bibfnamefont {Y.}~\bibnamefont {Zhang}}, \bibinfo {author} {\bibfnamefont
  {S.~V.}\ \bibnamefont {Dubonos}}, \bibinfo {author} {\bibfnamefont {I.~V.}\
  \bibnamefont {Grigorieva}}, \ and\ \bibinfo {author} {\bibfnamefont {A.~A.}\
  \bibnamefont {Firsov}},\ }\href@noop {} {\bibfield  {journal} {\bibinfo
  {journal} {science}\ }\textbf {\bibinfo {volume} {306}},\ \bibinfo {pages}
  {666} (\bibinfo {year} {2004})}\BibitemShut {NoStop}%
\bibitem [{\citenamefont {Mas-Ballesté}\ \emph {et~al.}(2011)\citenamefont
  {Mas-Ballesté}, \citenamefont {Gómez-Navarro}, \citenamefont
  {Gómez-Herrero},\ and\ \citenamefont {Zamora}}]{C0NR00323A}%
  \BibitemOpen
  \bibfield  {author} {\bibinfo {author} {\bibfnamefont {R.}~\bibnamefont
  {Mas-Ballesté}}, \bibinfo {author} {\bibfnamefont {C.}~\bibnamefont
  {Gómez-Navarro}}, \bibinfo {author} {\bibfnamefont {J.}~\bibnamefont
  {Gómez-Herrero}}, \ and\ \bibinfo {author} {\bibfnamefont {F.}~\bibnamefont
  {Zamora}},\ }\href {\doibase 10.1039/C0NR00323A} {\bibfield  {journal}
  {\bibinfo  {journal} {Nanoscale}\ }\textbf {\bibinfo {volume} {3}},\ \bibinfo
  {pages} {20} (\bibinfo {year} {2011})}\BibitemShut {NoStop}%
\bibitem [{\citenamefont {Balendhran}\ \emph {et~al.}(2015)\citenamefont
  {Balendhran}, \citenamefont {Walia}, \citenamefont {Nili}, \citenamefont
  {Sriram},\ and\ \citenamefont {Bhaskaran}}]{doi:10.1002/smll.201402041}%
  \BibitemOpen
  \bibfield  {author} {\bibinfo {author} {\bibfnamefont {S.}~\bibnamefont
  {Balendhran}}, \bibinfo {author} {\bibfnamefont {S.}~\bibnamefont {Walia}},
  \bibinfo {author} {\bibfnamefont {H.}~\bibnamefont {Nili}}, \bibinfo {author}
  {\bibfnamefont {S.}~\bibnamefont {Sriram}}, \ and\ \bibinfo {author}
  {\bibfnamefont {M.}~\bibnamefont {Bhaskaran}},\ }\href {\doibase
  10.1002/smll.201402041} {\bibfield  {journal} {\bibinfo  {journal} {Small}\
  }\textbf {\bibinfo {volume} {11}},\ \bibinfo {pages} {640} (\bibinfo {year}
  {2015})}\BibitemShut {NoStop}%
\bibitem [{\citenamefont {Xu}\ \emph {et~al.}(2013)\citenamefont {Xu},
  \citenamefont {Liang}, \citenamefont {Shi},\ and\ \citenamefont
  {Chen}}]{doi:10.1021/cr300263a}%
  \BibitemOpen
  \bibfield  {author} {\bibinfo {author} {\bibfnamefont {M.}~\bibnamefont
  {Xu}}, \bibinfo {author} {\bibfnamefont {T.}~\bibnamefont {Liang}}, \bibinfo
  {author} {\bibfnamefont {M.}~\bibnamefont {Shi}}, \ and\ \bibinfo {author}
  {\bibfnamefont {H.}~\bibnamefont {Chen}},\ }\href {\doibase
  10.1021/cr300263a} {\bibfield  {journal} {\bibinfo  {journal} {Chemical
  Reviews}\ }\textbf {\bibinfo {volume} {113}},\ \bibinfo {pages} {3766}
  (\bibinfo {year} {2013})},\ \bibinfo {note} {pMID: 23286380}\BibitemShut
  {NoStop}%
\bibitem [{\citenamefont {Persson}\ \emph {et~al.}(2019)\citenamefont
  {Persson}, \citenamefont {Halim}, \citenamefont {Lind}, \citenamefont
  {Hansen}, \citenamefont {Wagner}, \citenamefont {Naslund}, \citenamefont
  {Darakchieva}, \citenamefont {Palisaitis}, \citenamefont {Rosen},\ and\
  \citenamefont {Persson}}]{doi:10.1002/adma.201805472}%
  \BibitemOpen
  \bibfield  {author} {\bibinfo {author} {\bibfnamefont {I.}~\bibnamefont
  {Persson}}, \bibinfo {author} {\bibfnamefont {J.}~\bibnamefont {Halim}},
  \bibinfo {author} {\bibfnamefont {H.}~\bibnamefont {Lind}}, \bibinfo {author}
  {\bibfnamefont {T.~W.}\ \bibnamefont {Hansen}}, \bibinfo {author}
  {\bibfnamefont {J.~B.}\ \bibnamefont {Wagner}}, \bibinfo {author}
  {\bibfnamefont {L.-A.}\ \bibnamefont {Naslund}}, \bibinfo {author}
  {\bibfnamefont {V.}~\bibnamefont {Darakchieva}}, \bibinfo {author}
  {\bibfnamefont {J.}~\bibnamefont {Palisaitis}}, \bibinfo {author}
  {\bibfnamefont {J.}~\bibnamefont {Rosen}}, \ and\ \bibinfo {author}
  {\bibfnamefont {P.~O.~A.}\ \bibnamefont {Persson}},\ }\href {\doibase
  10.1002/adma.201805472} {\bibfield  {journal} {\bibinfo  {journal} {Advanced
  Materials}\ }\textbf {\bibinfo {volume} {31}},\ \bibinfo {pages} {1805472}
  (\bibinfo {year} {2019})}\BibitemShut {NoStop}%
\bibitem [{\citenamefont {Miao}\ \emph {et~al.}(2017)\citenamefont {Miao},
  \citenamefont {Xu}, \citenamefont {Bristowe}, \citenamefont {Zhou},\ and\
  \citenamefont {Sun}}]{doi:10.1021/jacs.7b05133}%
  \BibitemOpen
  \bibfield  {author} {\bibinfo {author} {\bibfnamefont {N.}~\bibnamefont
  {Miao}}, \bibinfo {author} {\bibfnamefont {B.}~\bibnamefont {Xu}}, \bibinfo
  {author} {\bibfnamefont {N.~C.}\ \bibnamefont {Bristowe}}, \bibinfo {author}
  {\bibfnamefont {J.}~\bibnamefont {Zhou}}, \ and\ \bibinfo {author}
  {\bibfnamefont {Z.}~\bibnamefont {Sun}},\ }\href {\doibase
  10.1021/jacs.7b05133} {\bibfield  {journal} {\bibinfo  {journal} {Journal of
  the American Chemical Society}\ }\textbf {\bibinfo {volume} {139}},\ \bibinfo
  {pages} {11125} (\bibinfo {year} {2017})},\ \bibinfo {note} {pMID:
  28731338},\ \Eprint
  {http://arxiv.org/abs/https://doi.org/10.1021/jacs.7b05133}
  {https://doi.org/10.1021/jacs.7b05133} \BibitemShut {NoStop}%
\bibitem [{\citenamefont {Yi}\ \emph {et~al.}(2019)\citenamefont {Yi},
  \citenamefont {Chen}, \citenamefont {Wang}, \citenamefont {Ding},
  \citenamefont {Yang},\ and\ \citenamefont {Liu}}]{C9TC02030F}%
  \BibitemOpen
  \bibfield  {author} {\bibinfo {author} {\bibfnamefont {W.}~\bibnamefont
  {Yi}}, \bibinfo {author} {\bibfnamefont {X.}~\bibnamefont {Chen}}, \bibinfo
  {author} {\bibfnamefont {Z.}~\bibnamefont {Wang}}, \bibinfo {author}
  {\bibfnamefont {Y.}~\bibnamefont {Ding}}, \bibinfo {author} {\bibfnamefont
  {B.}~\bibnamefont {Yang}}, \ and\ \bibinfo {author} {\bibfnamefont
  {X.}~\bibnamefont {Liu}},\ }\href {\doibase 10.1039/C9TC02030F} {\bibfield
  {journal} {\bibinfo  {journal} {J. Mater. Chem. C}\ ,\ } (\bibinfo {year}
  {2019})}\BibitemShut {NoStop}%
\bibitem [{\citenamefont {Shojaei}\ and\ \citenamefont
  {Kang}(2017)}]{C7TC02346D}%
  \BibitemOpen
  \bibfield  {author} {\bibinfo {author} {\bibfnamefont {F.}~\bibnamefont
  {Shojaei}}\ and\ \bibinfo {author} {\bibfnamefont {H.~S.}\ \bibnamefont
  {Kang}},\ }\href {\doibase 10.1039/C7TC02346D} {\bibfield  {journal}
  {\bibinfo  {journal} {J. Mater. Chem. C}\ }\textbf {\bibinfo {volume} {5}},\
  \bibinfo {pages} {11267} (\bibinfo {year} {2017})}\BibitemShut {NoStop}%
\bibitem [{\citenamefont {Sun}\ \emph {et~al.}(2018)\citenamefont {Sun},
  \citenamefont {Meng}, \citenamefont {Wang}, \citenamefont {Wang},\ and\
  \citenamefont {Ni}}]{C8TA02494D}%
  \BibitemOpen
  \bibfield  {author} {\bibinfo {author} {\bibfnamefont {S.}~\bibnamefont
  {Sun}}, \bibinfo {author} {\bibfnamefont {F.}~\bibnamefont {Meng}}, \bibinfo
  {author} {\bibfnamefont {H.}~\bibnamefont {Wang}}, \bibinfo {author}
  {\bibfnamefont {H.}~\bibnamefont {Wang}}, \ and\ \bibinfo {author}
  {\bibfnamefont {Y.}~\bibnamefont {Ni}},\ }\href {\doibase 10.1039/C8TA02494D}
  {\bibfield  {journal} {\bibinfo  {journal} {J. Mater. Chem. A}\ }\textbf
  {\bibinfo {volume} {6}},\ \bibinfo {pages} {11890} (\bibinfo {year}
  {2018})}\BibitemShut {NoStop}%
\bibitem [{\citenamefont {Ghosh}\ \emph {et~al.}(2018)\citenamefont {Ghosh},
  \citenamefont {Puri}, \citenamefont {Agarwal},\ and\ \citenamefont
  {Bhowmick}}]{doi:10.1021/acs.jpcc.8b06668}%
  \BibitemOpen
  \bibfield  {author} {\bibinfo {author} {\bibfnamefont {B.}~\bibnamefont
  {Ghosh}}, \bibinfo {author} {\bibfnamefont {S.}~\bibnamefont {Puri}},
  \bibinfo {author} {\bibfnamefont {A.}~\bibnamefont {Agarwal}}, \ and\
  \bibinfo {author} {\bibfnamefont {S.}~\bibnamefont {Bhowmick}},\ }\href
  {\doibase 10.1021/acs.jpcc.8b06668} {\bibfield  {journal} {\bibinfo
  {journal} {The Journal of Physical Chemistry C}\ }\textbf {\bibinfo {volume}
  {122}},\ \bibinfo {pages} {18185} (\bibinfo {year} {2018})},\ \Eprint
  {http://arxiv.org/abs/https://doi.org/10.1021/acs.jpcc.8b06668}
  {https://doi.org/10.1021/acs.jpcc.8b06668} \BibitemShut {NoStop}%
\bibitem [{\citenamefont {Lu}\ \emph {et~al.}(2018)\citenamefont {Lu},
  \citenamefont {Zhuo}, \citenamefont {Guo}, \citenamefont {Wu}, \citenamefont
  {Fa}, \citenamefont {Wu},\ and\ \citenamefont
  {Zeng}}]{doi:10.1021/acs.jpclett.8b00595}%
  \BibitemOpen
  \bibfield  {author} {\bibinfo {author} {\bibfnamefont {N.}~\bibnamefont
  {Lu}}, \bibinfo {author} {\bibfnamefont {Z.}~\bibnamefont {Zhuo}}, \bibinfo
  {author} {\bibfnamefont {H.}~\bibnamefont {Guo}}, \bibinfo {author}
  {\bibfnamefont {P.}~\bibnamefont {Wu}}, \bibinfo {author} {\bibfnamefont
  {W.}~\bibnamefont {Fa}}, \bibinfo {author} {\bibfnamefont {X.}~\bibnamefont
  {Wu}}, \ and\ \bibinfo {author} {\bibfnamefont {X.~C.}\ \bibnamefont
  {Zeng}},\ }\href {\doibase 10.1021/acs.jpclett.8b00595} {\bibfield  {journal}
  {\bibinfo  {journal} {The Journal of Physical Chemistry Letters}\ }\textbf
  {\bibinfo {volume} {9}},\ \bibinfo {pages} {1728} (\bibinfo {year} {2018})},\
  \bibinfo {note} {pMID: 29558132},\ \Eprint
  {http://arxiv.org/abs/https://doi.org/10.1021/acs.jpclett.8b00595}
  {https://doi.org/10.1021/acs.jpclett.8b00595} \BibitemShut {NoStop}%
\bibitem [{\citenamefont {Liu}\ \emph {et~al.}(2018{\natexlab{a}})\citenamefont
  {Liu}, \citenamefont {Liu}, \citenamefont {Ye},\ and\ \citenamefont
  {Yan}}]{C7TA10248H}%
  \BibitemOpen
  \bibfield  {author} {\bibinfo {author} {\bibfnamefont {J.}~\bibnamefont
  {Liu}}, \bibinfo {author} {\bibfnamefont {C.-S.}\ \bibnamefont {Liu}},
  \bibinfo {author} {\bibfnamefont {X.-J.}\ \bibnamefont {Ye}}, \ and\ \bibinfo
  {author} {\bibfnamefont {X.-H.}\ \bibnamefont {Yan}},\ }\href {\doibase
  10.1039/C7TA10248H} {\bibfield  {journal} {\bibinfo  {journal} {J. Mater.
  Chem. A}\ }\textbf {\bibinfo {volume} {6}},\ \bibinfo {pages} {3634}
  (\bibinfo {year} {2018}{\natexlab{a}})}\BibitemShut {NoStop}%
\bibitem [{\citenamefont {Liu}\ \emph {et~al.}(2018{\natexlab{b}})\citenamefont
  {Liu}, \citenamefont {Yang}, \citenamefont {Liu},\ and\ \citenamefont
  {Ye}}]{doi:10.1021/acsaem.8b00621}%
  \BibitemOpen
  \bibfield  {author} {\bibinfo {author} {\bibfnamefont {C.-S.}\ \bibnamefont
  {Liu}}, \bibinfo {author} {\bibfnamefont {X.-L.}\ \bibnamefont {Yang}},
  \bibinfo {author} {\bibfnamefont {J.}~\bibnamefont {Liu}}, \ and\ \bibinfo
  {author} {\bibfnamefont {X.-J.}\ \bibnamefont {Ye}},\ }\href {\doibase
  10.1021/acsaem.8b00621} {\bibfield  {journal} {\bibinfo  {journal} {ACS
  Applied Energy Materials}\ }\textbf {\bibinfo {volume} {1}},\ \bibinfo
  {pages} {3850} (\bibinfo {year} {2018}{\natexlab{b}})},\ \Eprint
  {http://arxiv.org/abs/https://doi.org/10.1021/acsaem.8b00621}
  {https://doi.org/10.1021/acsaem.8b00621} \BibitemShut {NoStop}%
\bibitem [{\citenamefont {Yao}\ \emph {et~al.}(2019)\citenamefont {Yao},
  \citenamefont {Zhang}, \citenamefont {Zhang}, \citenamefont {Chen},\ and\
  \citenamefont {Zhou}}]{YAO20195948}%
  \BibitemOpen
  \bibfield  {author} {\bibinfo {author} {\bibfnamefont {S.}~\bibnamefont
  {Yao}}, \bibinfo {author} {\bibfnamefont {X.}~\bibnamefont {Zhang}}, \bibinfo
  {author} {\bibfnamefont {Z.}~\bibnamefont {Zhang}}, \bibinfo {author}
  {\bibfnamefont {A.}~\bibnamefont {Chen}}, \ and\ \bibinfo {author}
  {\bibfnamefont {Z.}~\bibnamefont {Zhou}},\ }\href {\doibase
  https://doi.org/10.1016/j.ijhydene.2019.01.106} {\bibfield  {journal}
  {\bibinfo  {journal} {International Journal of Hydrogen Energy}\ }\textbf
  {\bibinfo {volume} {44}},\ \bibinfo {pages} {5948 } (\bibinfo {year}
  {2019})}\BibitemShut {NoStop}%
\bibitem [{\citenamefont {Donohue}\ and\ \citenamefont
  {Young}(1970)}]{DONOHUE1970143}%
  \BibitemOpen
  \bibfield  {author} {\bibinfo {author} {\bibfnamefont {P.}~\bibnamefont
  {Donohue}}\ and\ \bibinfo {author} {\bibfnamefont {H.}~\bibnamefont
  {Young}},\ }\href {\doibase https://doi.org/10.1016/0022-4596(70)90005-8}
  {\bibfield  {journal} {\bibinfo  {journal} {Journal of Solid State
  Chemistry}\ }\textbf {\bibinfo {volume} {1}},\ \bibinfo {pages} {143 }
  (\bibinfo {year} {1970})}\BibitemShut {NoStop}%
\bibitem [{\citenamefont {Gullman}\ and\ \citenamefont
  {Olofsson}(1972)}]{GULLMAN1972441}%
  \BibitemOpen
  \bibfield  {author} {\bibinfo {author} {\bibfnamefont {J.}~\bibnamefont
  {Gullman}}\ and\ \bibinfo {author} {\bibfnamefont {O.}~\bibnamefont
  {Olofsson}},\ }\href {\doibase https://doi.org/10.1016/0022-4596(72)90091-6}
  {\bibfield  {journal} {\bibinfo  {journal} {Journal of Solid State
  Chemistry}\ }\textbf {\bibinfo {volume} {5}},\ \bibinfo {pages} {441 }
  (\bibinfo {year} {1972})}\BibitemShut {NoStop}%
\bibitem [{\citenamefont {Jing}\ \emph
  {et~al.}(2017{\natexlab{a}})\citenamefont {Jing}, \citenamefont {Ma},
  \citenamefont {Li},\ and\ \citenamefont
  {Heine}}]{doi:10.1021/acs.nanolett.6b05143}%
  \BibitemOpen
  \bibfield  {author} {\bibinfo {author} {\bibfnamefont {Y.}~\bibnamefont
  {Jing}}, \bibinfo {author} {\bibfnamefont {Y.}~\bibnamefont {Ma}}, \bibinfo
  {author} {\bibfnamefont {Y.}~\bibnamefont {Li}}, \ and\ \bibinfo {author}
  {\bibfnamefont {T.}~\bibnamefont {Heine}},\ }\href {\doibase
  10.1021/acs.nanolett.6b05143} {\bibfield  {journal} {\bibinfo  {journal}
  {Nano Letters}\ }\textbf {\bibinfo {volume} {17}},\ \bibinfo {pages} {1833}
  (\bibinfo {year} {2017}{\natexlab{a}})},\ \bibinfo {note} {pMID: 28125237},\
  \Eprint {http://arxiv.org/abs/https://doi.org/10.1021/acs.nanolett.6b05143}
  {https://doi.org/10.1021/acs.nanolett.6b05143} \BibitemShut {NoStop}%
\bibitem [{\citenamefont {Li}\ \emph {et~al.}(2018)\citenamefont {Li},
  \citenamefont {Zhang}, \citenamefont {Li}, \citenamefont {Liang},\ and\
  \citenamefont {Zeng}}]{doi:10.1021/acsami.8b05655}%
  \BibitemOpen
  \bibfield  {author} {\bibinfo {author} {\bibfnamefont {P.}~\bibnamefont
  {Li}}, \bibinfo {author} {\bibfnamefont {W.}~\bibnamefont {Zhang}}, \bibinfo
  {author} {\bibfnamefont {D.}~\bibnamefont {Li}}, \bibinfo {author}
  {\bibfnamefont {C.}~\bibnamefont {Liang}}, \ and\ \bibinfo {author}
  {\bibfnamefont {X.~C.}\ \bibnamefont {Zeng}},\ }\href {\doibase
  10.1021/acsami.8b05655} {\bibfield  {journal} {\bibinfo  {journal} {ACS
  Applied Materials \& Interfaces}\ }\textbf {\bibinfo {volume} {10}},\
  \bibinfo {pages} {19897} (\bibinfo {year} {2018})},\ \bibinfo {note} {pMID:
  29792327},\ \Eprint
  {http://arxiv.org/abs/https://doi.org/10.1021/acsami.8b05655}
  {https://doi.org/10.1021/acsami.8b05655} \BibitemShut {NoStop}%
\bibitem [{\citenamefont {Zhang}\ \emph {et~al.}(2017)\citenamefont {Zhang},
  \citenamefont {Jiao}, \citenamefont {He}, \citenamefont {Ma}, \citenamefont
  {Kou}, \citenamefont {Liao}, \citenamefont {Bottle},\ and\ \citenamefont
  {Du}}]{C7CP04758D}%
  \BibitemOpen
  \bibfield  {author} {\bibinfo {author} {\bibfnamefont {C.}~\bibnamefont
  {Zhang}}, \bibinfo {author} {\bibfnamefont {Y.}~\bibnamefont {Jiao}},
  \bibinfo {author} {\bibfnamefont {T.}~\bibnamefont {He}}, \bibinfo {author}
  {\bibfnamefont {F.}~\bibnamefont {Ma}}, \bibinfo {author} {\bibfnamefont
  {L.}~\bibnamefont {Kou}}, \bibinfo {author} {\bibfnamefont {T.}~\bibnamefont
  {Liao}}, \bibinfo {author} {\bibfnamefont {S.}~\bibnamefont {Bottle}}, \ and\
  \bibinfo {author} {\bibfnamefont {A.}~\bibnamefont {Du}},\ }\href {\doibase
  10.1039/C7CP04758D} {\bibfield  {journal} {\bibinfo  {journal} {Phys. Chem.
  Chem. Phys.}\ }\textbf {\bibinfo {volume} {19}},\ \bibinfo {pages} {25886}
  (\bibinfo {year} {2017})}\BibitemShut {NoStop}%
\bibitem [{\citenamefont {Deng}\ \emph {et~al.}(2019)\citenamefont {Deng},
  \citenamefont {Chen}, \citenamefont {Huang}, \citenamefont {Xiao},\ and\
  \citenamefont {Du}}]{doi:10.1021/acs.jpcc.8b11574}%
  \BibitemOpen
  \bibfield  {author} {\bibinfo {author} {\bibfnamefont {X.}~\bibnamefont
  {Deng}}, \bibinfo {author} {\bibfnamefont {X.}~\bibnamefont {Chen}}, \bibinfo
  {author} {\bibfnamefont {Y.}~\bibnamefont {Huang}}, \bibinfo {author}
  {\bibfnamefont {B.}~\bibnamefont {Xiao}}, \ and\ \bibinfo {author}
  {\bibfnamefont {H.}~\bibnamefont {Du}},\ }\href {\doibase
  10.1021/acs.jpcc.8b11574} {\bibfield  {journal} {\bibinfo  {journal} {The
  Journal of Physical Chemistry C}\ }\textbf {\bibinfo {volume} {123}},\
  \bibinfo {pages} {4721} (\bibinfo {year} {2019})},\ \Eprint
  {http://arxiv.org/abs/https://doi.org/10.1021/acs.jpcc.8b11574}
  {https://doi.org/10.1021/acs.jpcc.8b11574} \BibitemShut {NoStop}%
\bibitem [{\citenamefont {Zeng}\ \emph {et~al.}(2018)\citenamefont {Zeng},
  \citenamefont {Long}, \citenamefont {Zhang}, \citenamefont {Dong},
  \citenamefont {Li}, \citenamefont {Yi},\ and\ \citenamefont
  {Duan}}]{Zeng_2018}%
  \BibitemOpen
  \bibfield  {author} {\bibinfo {author} {\bibfnamefont {B.}~\bibnamefont
  {Zeng}}, \bibinfo {author} {\bibfnamefont {M.}~\bibnamefont {Long}}, \bibinfo
  {author} {\bibfnamefont {X.}~\bibnamefont {Zhang}}, \bibinfo {author}
  {\bibfnamefont {Y.}~\bibnamefont {Dong}}, \bibinfo {author} {\bibfnamefont
  {M.}~\bibnamefont {Li}}, \bibinfo {author} {\bibfnamefont {Y.}~\bibnamefont
  {Yi}}, \ and\ \bibinfo {author} {\bibfnamefont {H.}~\bibnamefont {Duan}},\
  }\href {\doibase 10.1088/1361-6463/aac0a4} {\bibfield  {journal} {\bibinfo
  {journal} {Journal of Physics D: Applied Physics}\ }\textbf {\bibinfo
  {volume} {51}},\ \bibinfo {pages} {235302} (\bibinfo {year}
  {2018})}\BibitemShut {NoStop}%
\bibitem [{\citenamefont {Wang}\ \emph {et~al.}(2018)\citenamefont {Wang},
  \citenamefont {Li}, \citenamefont {Wang},\ and\ \citenamefont
  {Nie}}]{Wang2018}%
  \BibitemOpen
  \bibfield  {author} {\bibinfo {author} {\bibfnamefont {Q.}~\bibnamefont
  {Wang}}, \bibinfo {author} {\bibfnamefont {J.-W.}\ \bibnamefont {Li}},
  \bibinfo {author} {\bibfnamefont {B.}~\bibnamefont {Wang}}, \ and\ \bibinfo
  {author} {\bibfnamefont {Y.-H.}\ \bibnamefont {Nie}},\ }\href {\doibase
  10.1007/s11467-018-0750-x} {\bibfield  {journal} {\bibinfo  {journal}
  {Frontiers of Physics}\ }\textbf {\bibinfo {volume} {13}},\ \bibinfo {pages}
  {138501} (\bibinfo {year} {2018})}\BibitemShut {NoStop}%
\bibitem [{\citenamefont {Li}\ \emph {et~al.}(2019{\natexlab{a}})\citenamefont
  {Li}, \citenamefont {Huang}, \citenamefont {Ma}, \citenamefont {Zhu},
  \citenamefont {Li}, \citenamefont {Xia}, \citenamefont {Zeng},\ and\
  \citenamefont {Jia}}]{C8CP06310A}%
  \BibitemOpen
  \bibfield  {author} {\bibinfo {author} {\bibfnamefont {R.}~\bibnamefont
  {Li}}, \bibinfo {author} {\bibfnamefont {X.}~\bibnamefont {Huang}}, \bibinfo
  {author} {\bibfnamefont {X.}~\bibnamefont {Ma}}, \bibinfo {author}
  {\bibfnamefont {Z.}~\bibnamefont {Zhu}}, \bibinfo {author} {\bibfnamefont
  {C.}~\bibnamefont {Li}}, \bibinfo {author} {\bibfnamefont {C.}~\bibnamefont
  {Xia}}, \bibinfo {author} {\bibfnamefont {Z.}~\bibnamefont {Zeng}}, \ and\
  \bibinfo {author} {\bibfnamefont {Y.}~\bibnamefont {Jia}},\ }\href {\doibase
  10.1039/C8CP06310A} {\bibfield  {journal} {\bibinfo  {journal} {Phys. Chem.
  Chem. Phys.}\ }\textbf {\bibinfo {volume} {21}},\ \bibinfo {pages} {275}
  (\bibinfo {year} {2019}{\natexlab{a}})}\BibitemShut {NoStop}%
\bibitem [{\citenamefont {Niu}\ \emph {et~al.}(2019)\citenamefont {Niu},
  \citenamefont {Cai}, \citenamefont {Pang}, \citenamefont {Li}, \citenamefont
  {Zhang},\ and\ \citenamefont {Yang}}]{NIU201937}%
  \BibitemOpen
  \bibfield  {author} {\bibinfo {author} {\bibfnamefont {F.}~\bibnamefont
  {Niu}}, \bibinfo {author} {\bibfnamefont {M.}~\bibnamefont {Cai}}, \bibinfo
  {author} {\bibfnamefont {J.}~\bibnamefont {Pang}}, \bibinfo {author}
  {\bibfnamefont {X.}~\bibnamefont {Li}}, \bibinfo {author} {\bibfnamefont
  {G.}~\bibnamefont {Zhang}}, \ and\ \bibinfo {author} {\bibfnamefont
  {D.}~\bibnamefont {Yang}},\ }\href {\doibase
  https://doi.org/10.1016/j.susc.2019.02.008} {\bibfield  {journal} {\bibinfo
  {journal} {Surface Science}\ }\textbf {\bibinfo {volume} {684}},\ \bibinfo
  {pages} {37 } (\bibinfo {year} {2019})}\BibitemShut {NoStop}%
\bibitem [{\citenamefont {Tian}\ \emph {et~al.}(2019)\citenamefont {Tian},
  \citenamefont {Huang}, \citenamefont {Guo}, \citenamefont {Shu},
  \citenamefont {Wang},\ and\ \citenamefont {Dai}}]{TIAN2019181}%
  \BibitemOpen
  \bibfield  {author} {\bibinfo {author} {\bibfnamefont {B.}~\bibnamefont
  {Tian}}, \bibinfo {author} {\bibfnamefont {T.}~\bibnamefont {Huang}},
  \bibinfo {author} {\bibfnamefont {J.}~\bibnamefont {Guo}}, \bibinfo {author}
  {\bibfnamefont {H.}~\bibnamefont {Shu}}, \bibinfo {author} {\bibfnamefont
  {Y.}~\bibnamefont {Wang}}, \ and\ \bibinfo {author} {\bibfnamefont
  {J.}~\bibnamefont {Dai}},\ }\href {\doibase
  https://doi.org/10.1016/j.vacuum.2019.03.020} {\bibfield  {journal} {\bibinfo
   {journal} {Vacuum}\ }\textbf {\bibinfo {volume} {164}},\ \bibinfo {pages}
  {181 } (\bibinfo {year} {2019})}\BibitemShut {NoStop}%
\bibitem [{\citenamefont {Wu}\ \emph {et~al.}(2019)\citenamefont {Wu},
  \citenamefont {Huang}, \citenamefont {Zhong}, \citenamefont {Yu},
  \citenamefont {Zhang},\ and\ \citenamefont {Zeng}}]{C9NR03255J}%
  \BibitemOpen
  \bibfield  {author} {\bibinfo {author} {\bibfnamefont {H.-H.}\ \bibnamefont
  {Wu}}, \bibinfo {author} {\bibfnamefont {H.}~\bibnamefont {Huang}}, \bibinfo
  {author} {\bibfnamefont {J.}~\bibnamefont {Zhong}}, \bibinfo {author}
  {\bibfnamefont {S.}~\bibnamefont {Yu}}, \bibinfo {author} {\bibfnamefont
  {Q.}~\bibnamefont {Zhang}}, \ and\ \bibinfo {author} {\bibfnamefont {X.~C.}\
  \bibnamefont {Zeng}},\ }\href {\doibase 10.1039/C9NR03255J} {\bibfield
  {journal} {\bibinfo  {journal} {Nanoscale}\ ,\ } (\bibinfo {year}
  {2019})}\BibitemShut {NoStop}%
\bibitem [{\citenamefont {Zhang}\ \emph {et~al.}(2019)\citenamefont {Zhang},
  \citenamefont {Li}, \citenamefont {Fu}, \citenamefont {Zhao}, \citenamefont
  {Niu}, \citenamefont {Li}, \citenamefont {Zeng}, \citenamefont {Wang},
  \citenamefont {Xia},\ and\ \citenamefont {Jia}}]{zhang2019strong}%
  \BibitemOpen
  \bibfield  {author} {\bibinfo {author} {\bibfnamefont {S.}~\bibnamefont
  {Zhang}}, \bibinfo {author} {\bibfnamefont {R.}~\bibnamefont {Li}}, \bibinfo
  {author} {\bibfnamefont {X.}~\bibnamefont {Fu}}, \bibinfo {author}
  {\bibfnamefont {Y.}~\bibnamefont {Zhao}}, \bibinfo {author} {\bibfnamefont
  {C.}~\bibnamefont {Niu}}, \bibinfo {author} {\bibfnamefont {C.}~\bibnamefont
  {Li}}, \bibinfo {author} {\bibfnamefont {Z.}~\bibnamefont {Zeng}}, \bibinfo
  {author} {\bibfnamefont {S.}~\bibnamefont {Wang}}, \bibinfo {author}
  {\bibfnamefont {C.}~\bibnamefont {Xia}}, \ and\ \bibinfo {author}
  {\bibfnamefont {Y.}~\bibnamefont {Jia}},\ }\href@noop {} {\bibfield
  {journal} {\bibinfo  {journal} {Nanoscale research letters}\ }\textbf
  {\bibinfo {volume} {14}},\ \bibinfo {pages} {307} (\bibinfo {year}
  {2019})}\BibitemShut {NoStop}%
\bibitem [{\citenamefont {Gong}\ and\ \citenamefont
  {Zhang}(2019{\natexlab{a}})}]{gong2019two}%
  \BibitemOpen
  \bibfield  {author} {\bibinfo {author} {\bibfnamefont {C.}~\bibnamefont
  {Gong}}\ and\ \bibinfo {author} {\bibfnamefont {X.}~\bibnamefont {Zhang}},\
  }\href@noop {} {\bibfield  {journal} {\bibinfo  {journal} {Science}\ }\textbf
  {\bibinfo {volume} {363}},\ \bibinfo {pages} {eaav4450} (\bibinfo {year}
  {2019}{\natexlab{a}})}\BibitemShut {NoStop}%
\bibitem [{\citenamefont {Gong}\ \emph {et~al.}(2017)\citenamefont {Gong},
  \citenamefont {Li}, \citenamefont {Li}, \citenamefont {Ji}, \citenamefont
  {Stern}, \citenamefont {Xia}, \citenamefont {Cao}, \citenamefont {Bao},
  \citenamefont {Wang}, \citenamefont {Wang}, \citenamefont {Qiu},
  \citenamefont {Cava}, \citenamefont {Louie}, \citenamefont {Xia},\ and\
  \citenamefont {Zhang}}]{Gong2017}%
  \BibitemOpen
  \bibfield  {author} {\bibinfo {author} {\bibfnamefont {C.}~\bibnamefont
  {Gong}}, \bibinfo {author} {\bibfnamefont {L.}~\bibnamefont {Li}}, \bibinfo
  {author} {\bibfnamefont {Z.}~\bibnamefont {Li}}, \bibinfo {author}
  {\bibfnamefont {H.}~\bibnamefont {Ji}}, \bibinfo {author} {\bibfnamefont
  {A.}~\bibnamefont {Stern}}, \bibinfo {author} {\bibfnamefont
  {Y.}~\bibnamefont {Xia}}, \bibinfo {author} {\bibfnamefont {T.}~\bibnamefont
  {Cao}}, \bibinfo {author} {\bibfnamefont {W.}~\bibnamefont {Bao}}, \bibinfo
  {author} {\bibfnamefont {C.}~\bibnamefont {Wang}}, \bibinfo {author}
  {\bibfnamefont {Y.}~\bibnamefont {Wang}}, \bibinfo {author} {\bibfnamefont
  {Z.~Q.}\ \bibnamefont {Qiu}}, \bibinfo {author} {\bibfnamefont {R.~J.}\
  \bibnamefont {Cava}}, \bibinfo {author} {\bibfnamefont {S.~G.}\ \bibnamefont
  {Louie}}, \bibinfo {author} {\bibfnamefont {J.}~\bibnamefont {Xia}}, \ and\
  \bibinfo {author} {\bibfnamefont {X.}~\bibnamefont {Zhang}},\ }\href
  {https://doi.org/10.1038/nature22060} {\bibfield  {journal} {\bibinfo
  {journal} {Nature}\ }\textbf {\bibinfo {volume} {546}},\ \bibinfo {pages}
  {265 EP } (\bibinfo {year} {2017})}\BibitemShut {NoStop}%
\bibitem [{\citenamefont {Huang}\ \emph {et~al.}(2017)\citenamefont {Huang},
  \citenamefont {Clark}, \citenamefont {Navarro-Moratalla}, \citenamefont
  {Klein}, \citenamefont {Cheng}, \citenamefont {Seyler}, \citenamefont
  {Zhong}, \citenamefont {Schmidgall}, \citenamefont {McGuire}, \citenamefont
  {Cobden}, \citenamefont {Yao}, \citenamefont {Xiao}, \citenamefont
  {Jarillo-Herrero},\ and\ \citenamefont {Xu}}]{Huang2017}%
  \BibitemOpen
  \bibfield  {author} {\bibinfo {author} {\bibfnamefont {B.}~\bibnamefont
  {Huang}}, \bibinfo {author} {\bibfnamefont {G.}~\bibnamefont {Clark}},
  \bibinfo {author} {\bibfnamefont {E.}~\bibnamefont {Navarro-Moratalla}},
  \bibinfo {author} {\bibfnamefont {D.~R.}\ \bibnamefont {Klein}}, \bibinfo
  {author} {\bibfnamefont {R.}~\bibnamefont {Cheng}}, \bibinfo {author}
  {\bibfnamefont {K.~L.}\ \bibnamefont {Seyler}}, \bibinfo {author}
  {\bibfnamefont {D.}~\bibnamefont {Zhong}}, \bibinfo {author} {\bibfnamefont
  {E.}~\bibnamefont {Schmidgall}}, \bibinfo {author} {\bibfnamefont {M.~A.}\
  \bibnamefont {McGuire}}, \bibinfo {author} {\bibfnamefont {D.~H.}\
  \bibnamefont {Cobden}}, \bibinfo {author} {\bibfnamefont {W.}~\bibnamefont
  {Yao}}, \bibinfo {author} {\bibfnamefont {D.}~\bibnamefont {Xiao}}, \bibinfo
  {author} {\bibfnamefont {P.}~\bibnamefont {Jarillo-Herrero}}, \ and\ \bibinfo
  {author} {\bibfnamefont {X.}~\bibnamefont {Xu}},\ }\href
  {https://doi.org/10.1038/nature22391} {\bibfield  {journal} {\bibinfo
  {journal} {Nature}\ }\textbf {\bibinfo {volume} {546}},\ \bibinfo {pages}
  {270 EP } (\bibinfo {year} {2017})}\BibitemShut {NoStop}%
\bibitem [{\citenamefont {Gong}\ and\ \citenamefont
  {Zhang}(2019{\natexlab{b}})}]{Gongeaav4450}%
  \BibitemOpen
  \bibfield  {author} {\bibinfo {author} {\bibfnamefont {C.}~\bibnamefont
  {Gong}}\ and\ \bibinfo {author} {\bibfnamefont {X.}~\bibnamefont {Zhang}},\
  }\href {\doibase 10.1126/science.aav4450} {\bibfield  {journal} {\bibinfo
  {journal} {Science}\ }\textbf {\bibinfo {volume} {363}},\ \bibinfo {pages}
  {706} (\bibinfo {year} {2019}{\natexlab{b}})}\BibitemShut {NoStop}%
\bibitem [{\citenamefont {Gibertini}\ \emph {et~al.}(2019)\citenamefont
  {Gibertini}, \citenamefont {Koperski}, \citenamefont {Morpurgo},\ and\
  \citenamefont {Novoselov}}]{Gibertini2019}%
  \BibitemOpen
  \bibfield  {author} {\bibinfo {author} {\bibfnamefont {M.}~\bibnamefont
  {Gibertini}}, \bibinfo {author} {\bibfnamefont {M.}~\bibnamefont {Koperski}},
  \bibinfo {author} {\bibfnamefont {A.~F.}\ \bibnamefont {Morpurgo}}, \ and\
  \bibinfo {author} {\bibfnamefont {K.~S.}\ \bibnamefont {Novoselov}},\ }\href
  {\doibase 10.1038/s41565-019-0438-6} {\bibfield  {journal} {\bibinfo
  {journal} {Nature Nanotechnology}\ }\textbf {\bibinfo {volume} {14}},\
  \bibinfo {pages} {408} (\bibinfo {year} {2019})}\BibitemShut {NoStop}%
\bibitem [{\citenamefont {Jiang}\ \emph {et~al.}(2018)\citenamefont {Jiang},
  \citenamefont {Shan},\ and\ \citenamefont {Mak}}]{jiang2018electric}%
  \BibitemOpen
  \bibfield  {author} {\bibinfo {author} {\bibfnamefont {S.}~\bibnamefont
  {Jiang}}, \bibinfo {author} {\bibfnamefont {J.}~\bibnamefont {Shan}}, \ and\
  \bibinfo {author} {\bibfnamefont {K.~F.}\ \bibnamefont {Mak}},\ }\href@noop
  {} {\bibfield  {journal} {\bibinfo  {journal} {Nature materials}\ }\textbf
  {\bibinfo {volume} {17}},\ \bibinfo {pages} {406} (\bibinfo {year}
  {2018})}\BibitemShut {NoStop}%
\bibitem [{\citenamefont {Sivadas}\ \emph {et~al.}(2018)\citenamefont
  {Sivadas}, \citenamefont {Okamoto}, \citenamefont {Xu}, \citenamefont
  {Fennie},\ and\ \citenamefont {Xiao}}]{sivadas2018stacking}%
  \BibitemOpen
  \bibfield  {author} {\bibinfo {author} {\bibfnamefont {N.}~\bibnamefont
  {Sivadas}}, \bibinfo {author} {\bibfnamefont {S.}~\bibnamefont {Okamoto}},
  \bibinfo {author} {\bibfnamefont {X.}~\bibnamefont {Xu}}, \bibinfo {author}
  {\bibfnamefont {C.~J.}\ \bibnamefont {Fennie}}, \ and\ \bibinfo {author}
  {\bibfnamefont {D.}~\bibnamefont {Xiao}},\ }\href@noop {} {\bibfield
  {journal} {\bibinfo  {journal} {Nano letters}\ }\textbf {\bibinfo {volume}
  {18}},\ \bibinfo {pages} {7658} (\bibinfo {year} {2018})}\BibitemShut
  {NoStop}%
\bibitem [{\citenamefont {Jiang}\ \emph {et~al.}(2019)\citenamefont {Jiang},
  \citenamefont {Wang}, \citenamefont {Chen}, \citenamefont {Zhong},
  \citenamefont {Yuan}, \citenamefont {Lu},\ and\ \citenamefont
  {Ji}}]{jiang2019stacking}%
  \BibitemOpen
  \bibfield  {author} {\bibinfo {author} {\bibfnamefont {P.}~\bibnamefont
  {Jiang}}, \bibinfo {author} {\bibfnamefont {C.}~\bibnamefont {Wang}},
  \bibinfo {author} {\bibfnamefont {D.}~\bibnamefont {Chen}}, \bibinfo {author}
  {\bibfnamefont {Z.}~\bibnamefont {Zhong}}, \bibinfo {author} {\bibfnamefont
  {Z.}~\bibnamefont {Yuan}}, \bibinfo {author} {\bibfnamefont {Z.-Y.}\
  \bibnamefont {Lu}}, \ and\ \bibinfo {author} {\bibfnamefont {W.}~\bibnamefont
  {Ji}},\ }\href@noop {} {\bibfield  {journal} {\bibinfo  {journal} {Physical
  Review B}\ }\textbf {\bibinfo {volume} {99}},\ \bibinfo {pages} {144401}
  (\bibinfo {year} {2019})}\BibitemShut {NoStop}%
\bibitem [{\citenamefont {Morell}\ \emph {et~al.}(2019)\citenamefont {Morell},
  \citenamefont {Le{\'o}n}, \citenamefont {Miwa},\ and\ \citenamefont
  {Vargas}}]{morell2019control}%
  \BibitemOpen
  \bibfield  {author} {\bibinfo {author} {\bibfnamefont {E.~S.}\ \bibnamefont
  {Morell}}, \bibinfo {author} {\bibfnamefont {A.}~\bibnamefont {Le{\'o}n}},
  \bibinfo {author} {\bibfnamefont {R.~H.}\ \bibnamefont {Miwa}}, \ and\
  \bibinfo {author} {\bibfnamefont {P.}~\bibnamefont {Vargas}},\ }\href@noop {}
  {\bibfield  {journal} {\bibinfo  {journal} {2D Materials}\ }\textbf {\bibinfo
  {volume} {6}},\ \bibinfo {pages} {025020} (\bibinfo {year}
  {2019})}\BibitemShut {NoStop}%
\bibitem [{\citenamefont {Giannozzi{\it~et al.}}(2009)}]{espresso}%
  \BibitemOpen
  \bibfield  {author} {\bibinfo {author} {\bibfnamefont {P.}~\bibnamefont
  {Giannozzi{\it~et al.}}},\ }\href@noop {} {\bibfield  {journal} {\bibinfo
  {journal} {J. Phys.: Condens. Matter}\ }\textbf {\bibinfo {volume} {21}},\
  \bibinfo {pages} {395502} (\bibinfo {year} {2009})}\BibitemShut {NoStop}%
\bibitem [{\citenamefont {Perdew}\ \emph {et~al.}(1996)\citenamefont {Perdew},
  \citenamefont {Burke},\ and\ \citenamefont {Ernzerhof}}]{pbe}%
  \BibitemOpen
  \bibfield  {author} {\bibinfo {author} {\bibfnamefont {J.~P.}\ \bibnamefont
  {Perdew}}, \bibinfo {author} {\bibfnamefont {K.}~\bibnamefont {Burke}}, \
  and\ \bibinfo {author} {\bibfnamefont {M.}~\bibnamefont {Ernzerhof}},\
  }\href@noop {} {\bibfield  {journal} {\bibinfo  {journal} {Phys. Rev. Lett.}\
  }\textbf {\bibinfo {volume} {77}},\ \bibinfo {pages} {3865} (\bibinfo {year}
  {1996})}\BibitemShut {NoStop}%
\bibitem [{\citenamefont {Bl$\rm\ddot{u}$chl}(1994)}]{paw}%
  \BibitemOpen
  \bibfield  {author} {\bibinfo {author} {\bibfnamefont {P.~E.}\ \bibnamefont
  {Bl$\rm\ddot{u}$chl}},\ }\href@noop {} {\bibfield  {journal} {\bibinfo
  {journal} {Phys. Rev. B}\ }\textbf {\bibinfo {volume} {50}},\ \bibinfo
  {pages} {17953} (\bibinfo {year} {1994})}\BibitemShut {NoStop}%
\bibitem [{\citenamefont {Thonhauser}\ \emph {et~al.}(2015)\citenamefont
  {Thonhauser}, \citenamefont {Zuluaga}, \citenamefont {Arter}, \citenamefont
  {Berland}, \citenamefont {Schr{\"o}der},\ and\ \citenamefont
  {Hyldgaard}}]{thonhauserPRL2015}%
  \BibitemOpen
  \bibfield  {author} {\bibinfo {author} {\bibfnamefont {T.}~\bibnamefont
  {Thonhauser}}, \bibinfo {author} {\bibfnamefont {S.}~\bibnamefont {Zuluaga}},
  \bibinfo {author} {\bibfnamefont {C.}~\bibnamefont {Arter}}, \bibinfo
  {author} {\bibfnamefont {K.}~\bibnamefont {Berland}}, \bibinfo {author}
  {\bibfnamefont {E.}~\bibnamefont {Schr{\"o}der}}, \ and\ \bibinfo {author}
  {\bibfnamefont {P.}~\bibnamefont {Hyldgaard}},\ }\href@noop {} {\bibfield
  {journal} {\bibinfo  {journal} {Physical review letters}\ }\textbf {\bibinfo
  {volume} {115}},\ \bibinfo {pages} {136402} (\bibinfo {year}
  {2015})}\BibitemShut {NoStop}%
\bibitem [{\citenamefont {Thonhauser}\ \emph {et~al.}(2007)\citenamefont
  {Thonhauser}, \citenamefont {Cooper}, \citenamefont {Li}, \citenamefont
  {Puzder}, \citenamefont {Hyldgaard},\ and\ \citenamefont
  {Langreth}}]{thonhauserPRB2007}%
  \BibitemOpen
  \bibfield  {author} {\bibinfo {author} {\bibfnamefont {T.}~\bibnamefont
  {Thonhauser}}, \bibinfo {author} {\bibfnamefont {V.~R.}\ \bibnamefont
  {Cooper}}, \bibinfo {author} {\bibfnamefont {S.}~\bibnamefont {Li}}, \bibinfo
  {author} {\bibfnamefont {A.}~\bibnamefont {Puzder}}, \bibinfo {author}
  {\bibfnamefont {P.}~\bibnamefont {Hyldgaard}}, \ and\ \bibinfo {author}
  {\bibfnamefont {D.~C.}\ \bibnamefont {Langreth}},\ }\href@noop {} {\bibfield
  {journal} {\bibinfo  {journal} {Phys. Rev. B}\ }\textbf {\bibinfo {volume}
  {76}},\ \bibinfo {pages} {125112} (\bibinfo {year} {2007})}\BibitemShut
  {NoStop}%
\bibitem [{\citenamefont {Berland}\ \emph {et~al.}(2015)\citenamefont
  {Berland}, \citenamefont {Cooper}, \citenamefont {Lee}, \citenamefont
  {Schr{\"o}der}, \citenamefont {Thonhauser}, \citenamefont {Hyldgaard},\ and\
  \citenamefont {Lundqvist}}]{berlandRepProgPhys2015}%
  \BibitemOpen
  \bibfield  {author} {\bibinfo {author} {\bibfnamefont {K.}~\bibnamefont
  {Berland}}, \bibinfo {author} {\bibfnamefont {V.~R.}\ \bibnamefont {Cooper}},
  \bibinfo {author} {\bibfnamefont {K.}~\bibnamefont {Lee}}, \bibinfo {author}
  {\bibfnamefont {E.}~\bibnamefont {Schr{\"o}der}}, \bibinfo {author}
  {\bibfnamefont {T.}~\bibnamefont {Thonhauser}}, \bibinfo {author}
  {\bibfnamefont {P.}~\bibnamefont {Hyldgaard}}, \ and\ \bibinfo {author}
  {\bibfnamefont {B.~I.}\ \bibnamefont {Lundqvist}},\ }\href@noop {} {\bibfield
   {journal} {\bibinfo  {journal} {Reports on Progress in Physics}\ }\textbf
  {\bibinfo {volume} {78}},\ \bibinfo {pages} {066501} (\bibinfo {year}
  {2015})}\BibitemShut {NoStop}%
\bibitem [{\citenamefont {Langreth}\ \emph {et~al.}(2009)\citenamefont
  {Langreth}, \citenamefont {Lundqvist}, \citenamefont {Chakarova-K{\"a}ck},
  \citenamefont {Cooper}, \citenamefont {Dion}, \citenamefont {Hyldgaard},
  \citenamefont {Kelkkanen}, \citenamefont {Kleis}, \citenamefont {Kong},
  \citenamefont {Li} \emph {et~al.}}]{langrethJPhysC2009}%
  \BibitemOpen
  \bibfield  {author} {\bibinfo {author} {\bibfnamefont {D.}~\bibnamefont
  {Langreth}}, \bibinfo {author} {\bibfnamefont {B.~I.}\ \bibnamefont
  {Lundqvist}}, \bibinfo {author} {\bibfnamefont {S.~D.}\ \bibnamefont
  {Chakarova-K{\"a}ck}}, \bibinfo {author} {\bibfnamefont {V.}~\bibnamefont
  {Cooper}}, \bibinfo {author} {\bibfnamefont {M.}~\bibnamefont {Dion}},
  \bibinfo {author} {\bibfnamefont {P.}~\bibnamefont {Hyldgaard}}, \bibinfo
  {author} {\bibfnamefont {A.}~\bibnamefont {Kelkkanen}}, \bibinfo {author}
  {\bibfnamefont {J.}~\bibnamefont {Kleis}}, \bibinfo {author} {\bibfnamefont
  {L.}~\bibnamefont {Kong}}, \bibinfo {author} {\bibfnamefont {S.}~\bibnamefont
  {Li}},  \emph {et~al.},\ }\href@noop {} {\bibfield  {journal} {\bibinfo
  {journal} {Journal of Physics: Condensed Matter}\ }\textbf {\bibinfo {volume}
  {21}},\ \bibinfo {pages} {084203} (\bibinfo {year} {2009})}\BibitemShut
  {NoStop}%
\bibitem [{\citenamefont {Monkhorst}\ and\ \citenamefont {Pack}(1976)}]{mp}%
  \BibitemOpen
  \bibfield  {author} {\bibinfo {author} {\bibfnamefont {H.~J.}\ \bibnamefont
  {Monkhorst}}\ and\ \bibinfo {author} {\bibfnamefont {J.~D.}\ \bibnamefont
  {Pack}},\ }\href@noop {} {\bibfield  {journal} {\bibinfo  {journal} {Phys.
  Rev. B}\ }\textbf {\bibinfo {volume} {13}},\ \bibinfo {pages} {5188}
  (\bibinfo {year} {1976})}\BibitemShut {NoStop}%
\bibitem [{\citenamefont {Jing}\ \emph
  {et~al.}(2017{\natexlab{b}})\citenamefont {Jing}, \citenamefont {Ma},
  \citenamefont {Li},\ and\ \citenamefont {Heine}}]{jinggep3}%
  \BibitemOpen
  \bibfield  {author} {\bibinfo {author} {\bibfnamefont {Y.}~\bibnamefont
  {Jing}}, \bibinfo {author} {\bibfnamefont {Y.}~\bibnamefont {Ma}}, \bibinfo
  {author} {\bibfnamefont {Y.}~\bibnamefont {Li}}, \ and\ \bibinfo {author}
  {\bibfnamefont {T.}~\bibnamefont {Heine}},\ }\href {\doibase
  10.1021/acs.nanolett.6b05143} {\bibfield  {journal} {\bibinfo  {journal}
  {Nano Letters}\ }\textbf {\bibinfo {volume} {17}},\ \bibinfo {pages} {1833}
  (\bibinfo {year} {2017}{\natexlab{b}})},\ \bibinfo {note} {pMID: 28125237},\
  \Eprint {http://arxiv.org/abs/https://doi.org/10.1021/acs.nanolett.6b05143}
  {https://doi.org/10.1021/acs.nanolett.6b05143} \BibitemShut {NoStop}%
\bibitem [{\citenamefont {Kim}\ \emph {et~al.}(2018)\citenamefont {Kim},
  \citenamefont {Zhang}, \citenamefont {Lim}, \citenamefont {Lee},
  \citenamefont {Cho}, \citenamefont {Cho},\ and\ \citenamefont
  {Kang}}]{KIM2018126}%
  \BibitemOpen
  \bibfield  {author} {\bibinfo {author} {\bibfnamefont {D.}~\bibnamefont
  {Kim}}, \bibinfo {author} {\bibfnamefont {K.}~\bibnamefont {Zhang}}, \bibinfo
  {author} {\bibfnamefont {J.-M.}\ \bibnamefont {Lim}}, \bibinfo {author}
  {\bibfnamefont {G.-H.}\ \bibnamefont {Lee}}, \bibinfo {author} {\bibfnamefont
  {K.}~\bibnamefont {Cho}}, \bibinfo {author} {\bibfnamefont {M.}~\bibnamefont
  {Cho}}, \ and\ \bibinfo {author} {\bibfnamefont {Y.-M.}\ \bibnamefont
  {Kang}},\ }\href {\doibase https://doi.org/10.1016/j.mtener.2018.05.005}
  {\bibfield  {journal} {\bibinfo  {journal} {Materials Today Energy}\ }\textbf
  {\bibinfo {volume} {9}},\ \bibinfo {pages} {126 } (\bibinfo {year}
  {2018})}\BibitemShut {NoStop}%
\bibitem [{\citenamefont {Wang}\ \emph {et~al.}(2015)\citenamefont {Wang},
  \citenamefont {Yuan}, \citenamefont {Hong}, \citenamefont {Li},\ and\
  \citenamefont {Cui}}]{wang2015physical}%
  \BibitemOpen
  \bibfield  {author} {\bibinfo {author} {\bibfnamefont {H.}~\bibnamefont
  {Wang}}, \bibinfo {author} {\bibfnamefont {H.}~\bibnamefont {Yuan}}, \bibinfo
  {author} {\bibfnamefont {S.~S.}\ \bibnamefont {Hong}}, \bibinfo {author}
  {\bibfnamefont {Y.}~\bibnamefont {Li}}, \ and\ \bibinfo {author}
  {\bibfnamefont {Y.}~\bibnamefont {Cui}},\ }\href@noop {} {\bibfield
  {journal} {\bibinfo  {journal} {Chemical Society Reviews}\ }\textbf {\bibinfo
  {volume} {44}},\ \bibinfo {pages} {2664} (\bibinfo {year}
  {2015})}\BibitemShut {NoStop}%
\bibitem [{\citenamefont {Gong}\ \emph {et~al.}(2018)\citenamefont {Gong},
  \citenamefont {Yuan}, \citenamefont {Wu}, \citenamefont {Tang}, \citenamefont
  {Yang}, \citenamefont {Yang}, \citenamefont {Li}, \citenamefont {Liu},
  \citenamefont {van~de Groep}, \citenamefont {Brongersma} \emph
  {et~al.}}]{gong2018spatially}%
  \BibitemOpen
  \bibfield  {author} {\bibinfo {author} {\bibfnamefont {Y.}~\bibnamefont
  {Gong}}, \bibinfo {author} {\bibfnamefont {H.}~\bibnamefont {Yuan}}, \bibinfo
  {author} {\bibfnamefont {C.-L.}\ \bibnamefont {Wu}}, \bibinfo {author}
  {\bibfnamefont {P.}~\bibnamefont {Tang}}, \bibinfo {author} {\bibfnamefont
  {S.-Z.}\ \bibnamefont {Yang}}, \bibinfo {author} {\bibfnamefont
  {A.}~\bibnamefont {Yang}}, \bibinfo {author} {\bibfnamefont {G.}~\bibnamefont
  {Li}}, \bibinfo {author} {\bibfnamefont {B.}~\bibnamefont {Liu}}, \bibinfo
  {author} {\bibfnamefont {J.}~\bibnamefont {van~de Groep}}, \bibinfo {author}
  {\bibfnamefont {M.~L.}\ \bibnamefont {Brongersma}},  \emph {et~al.},\
  }\href@noop {} {\bibfield  {journal} {\bibinfo  {journal} {Nature
  nanotechnology}\ }\textbf {\bibinfo {volume} {13}},\ \bibinfo {pages} {294}
  (\bibinfo {year} {2018})}\BibitemShut {NoStop}%
\bibitem [{\citenamefont {Miwa}\ \emph {et~al.}(2015)\citenamefont {Miwa},
  \citenamefont {Venezuela},\ and\ \citenamefont {Morell}}]{miwa2015periodic}%
  \BibitemOpen
  \bibfield  {author} {\bibinfo {author} {\bibfnamefont {R.}~\bibnamefont
  {Miwa}}, \bibinfo {author} {\bibfnamefont {P.}~\bibnamefont {Venezuela}}, \
  and\ \bibinfo {author} {\bibfnamefont {E.~S.}\ \bibnamefont {Morell}},\
  }\href@noop {} {\bibfield  {journal} {\bibinfo  {journal} {Physical Review
  B}\ }\textbf {\bibinfo {volume} {92}},\ \bibinfo {pages} {115419} (\bibinfo
  {year} {2015})}\BibitemShut {NoStop}%
\bibitem [{\citenamefont {Zhao}\ \emph {et~al.}(2020)\citenamefont {Zhao},
  \citenamefont {Song}, \citenamefont {Wang}, \citenamefont {Riis-Jensen},
  \citenamefont {Fu}, \citenamefont {Deng}, \citenamefont {Wan}, \citenamefont
  {Kang}, \citenamefont {Ning}, \citenamefont {Dan} \emph
  {et~al.}}]{zhao2020engineering}%
  \BibitemOpen
  \bibfield  {author} {\bibinfo {author} {\bibfnamefont {X.}~\bibnamefont
  {Zhao}}, \bibinfo {author} {\bibfnamefont {P.}~\bibnamefont {Song}}, \bibinfo
  {author} {\bibfnamefont {C.}~\bibnamefont {Wang}}, \bibinfo {author}
  {\bibfnamefont {A.~C.}\ \bibnamefont {Riis-Jensen}}, \bibinfo {author}
  {\bibfnamefont {W.}~\bibnamefont {Fu}}, \bibinfo {author} {\bibfnamefont
  {Y.}~\bibnamefont {Deng}}, \bibinfo {author} {\bibfnamefont {D.}~\bibnamefont
  {Wan}}, \bibinfo {author} {\bibfnamefont {L.}~\bibnamefont {Kang}}, \bibinfo
  {author} {\bibfnamefont {S.}~\bibnamefont {Ning}}, \bibinfo {author}
  {\bibfnamefont {J.}~\bibnamefont {Dan}},  \emph {et~al.},\ }\href@noop {}
  {\bibfield  {journal} {\bibinfo  {journal} {Nature}\ }\textbf {\bibinfo
  {volume} {581}},\ \bibinfo {pages} {171} (\bibinfo {year}
  {2020})}\BibitemShut {NoStop}%
\bibitem [{\citenamefont {Qiu}\ \emph {et~al.}(2019)\citenamefont {Qiu},
  \citenamefont {Zhang}, \citenamefont {Lu},\ and\ \citenamefont
  {Liu}}]{qiuJPhysChemC2019}%
  \BibitemOpen
  \bibfield  {author} {\bibinfo {author} {\bibfnamefont {X.-L.}\ \bibnamefont
  {Qiu}}, \bibinfo {author} {\bibfnamefont {J.-F.}\ \bibnamefont {Zhang}},
  \bibinfo {author} {\bibfnamefont {Z.-Y.}\ \bibnamefont {Lu}}, \ and\ \bibinfo
  {author} {\bibfnamefont {K.}~\bibnamefont {Liu}},\ }\href@noop {} {\bibfield
  {journal} {\bibinfo  {journal} {The Journal of Physical Chemistry C}\
  }\textbf {\bibinfo {volume} {123}},\ \bibinfo {pages} {24698} (\bibinfo
  {year} {2019})}\BibitemShut {NoStop}%
\bibitem [{\citenamefont {Li}\ \emph {et~al.}(2019{\natexlab{b}})\citenamefont
  {Li}, \citenamefont {Jiang}, \citenamefont {Sivadas}, \citenamefont {Wang},
  \citenamefont {Xu}, \citenamefont {Weber}, \citenamefont {Goldberger},
  \citenamefont {Watanabe}, \citenamefont {Taniguchi}, \citenamefont {Fennie}
  \emph {et~al.}}]{li2019pressure}%
  \BibitemOpen
  \bibfield  {author} {\bibinfo {author} {\bibfnamefont {T.}~\bibnamefont
  {Li}}, \bibinfo {author} {\bibfnamefont {S.}~\bibnamefont {Jiang}}, \bibinfo
  {author} {\bibfnamefont {N.}~\bibnamefont {Sivadas}}, \bibinfo {author}
  {\bibfnamefont {Z.}~\bibnamefont {Wang}}, \bibinfo {author} {\bibfnamefont
  {Y.}~\bibnamefont {Xu}}, \bibinfo {author} {\bibfnamefont {D.}~\bibnamefont
  {Weber}}, \bibinfo {author} {\bibfnamefont {J.~E.}\ \bibnamefont
  {Goldberger}}, \bibinfo {author} {\bibfnamefont {K.}~\bibnamefont
  {Watanabe}}, \bibinfo {author} {\bibfnamefont {T.}~\bibnamefont {Taniguchi}},
  \bibinfo {author} {\bibfnamefont {C.~J.}\ \bibnamefont {Fennie}},  \emph
  {et~al.},\ }\href@noop {} {\bibfield  {journal} {\bibinfo  {journal} {Nature
  materials}\ }\textbf {\bibinfo {volume} {18}},\ \bibinfo {pages} {1303}
  (\bibinfo {year} {2019}{\natexlab{b}})}\BibitemShut {NoStop}%
\bibitem [{\citenamefont {Song}\ \emph {et~al.}(2019)\citenamefont {Song},
  \citenamefont {Fei}, \citenamefont {Yankowitz}, \citenamefont {Lin},
  \citenamefont {Jiang}, \citenamefont {Hwangbo}, \citenamefont {Zhang},
  \citenamefont {Sun}, \citenamefont {Taniguchi}, \citenamefont {Watanabe}
  \emph {et~al.}}]{song2019switching}%
  \BibitemOpen
  \bibfield  {author} {\bibinfo {author} {\bibfnamefont {T.}~\bibnamefont
  {Song}}, \bibinfo {author} {\bibfnamefont {Z.}~\bibnamefont {Fei}}, \bibinfo
  {author} {\bibfnamefont {M.}~\bibnamefont {Yankowitz}}, \bibinfo {author}
  {\bibfnamefont {Z.}~\bibnamefont {Lin}}, \bibinfo {author} {\bibfnamefont
  {Q.}~\bibnamefont {Jiang}}, \bibinfo {author} {\bibfnamefont
  {K.}~\bibnamefont {Hwangbo}}, \bibinfo {author} {\bibfnamefont
  {Q.}~\bibnamefont {Zhang}}, \bibinfo {author} {\bibfnamefont
  {B.}~\bibnamefont {Sun}}, \bibinfo {author} {\bibfnamefont {T.}~\bibnamefont
  {Taniguchi}}, \bibinfo {author} {\bibfnamefont {K.}~\bibnamefont {Watanabe}},
   \emph {et~al.},\ }\href@noop {} {\bibfield  {journal} {\bibinfo  {journal}
  {Nature materials}\ ,\ \bibinfo {pages} {1298}} (\bibinfo {year}
  {2019})}\BibitemShut {NoStop}%
\end{thebibliography}%

\end{document}